\documentclass[11pt,a4paper]{article}%
\usepackage{jheppub}
\usepackage{amsmath,amssymb}
\usepackage{graphicx}
\usepackage{epstopdf}
\usepackage{amsmath}
\usepackage{amsfonts}
\usepackage{amssymb}%
\providecommand{\U}[1]{\protect\rule{.1in}{.1in}}
\affiliation[a]{Department of Electrophysics, National Chiao Tung University, Hsinchu, ROC}
\affiliation[b]{Institute of Physics, National Chiao Tung University, Hsinchu, ROC}
\emailAdd{yiyang@mail.nctu.edu.tw}
\emailAdd{phy.pro.phy@gmail.com}
\abstract{We study confinement-deconfinement phase transition for heavy quarks in a
soft wall holographic QCD model. We consider a black hole background in an
Einstein-Maxwell-scalar system and add probe open strings to the background.
Combining the various configurations of the open strings and the phase
structure of the black hole background itself, we obtain the
confinement-deconfinement phase diagram for heavy quarks in the holographic
QCD model.}
\begin{document}

\title{Confinement-Deconfinement Phase Transition for Heavy Quarks in a Soft Wall
Holographic QCD Model}
\author{Yi Yang and Pei-Hung Yuan}
\maketitle

%

\setcounter{equation}{0}
\renewcommand{\theequation}{\arabic{section}.\arabic{equation}}%

\section{Introduction}

Confinement-deconfinement phase transition is an important and challenging
problem in QCD. Near the phase transition region, the interaction becomes very
strong so that the conventional perturbation method of QFT does not work. For
a long time, lattice QCD has been the only method to study strong interacted
QCD. Although lattice QCD works well for zero density, it encounters the sign
problem when considering finite quark density. See \cite{1009.4089,1203.5320}
for a review of the current status of lattice QCD. Recently, using the idea of
AdS/CFT duality from string theory, one is able to study QCD in the strongly
coupled region by studying its weakly coupled dual gravitational theory, i.e.
holographic QCD
\cite{0306018,0311270,0304032,0611099,0412141,0507073,0501128,0602229,0801.4383,0806.3830,0804.0434,1005.4690,1006.5461,1012.1864,1103.5389,1108.2029,1209.4512}%
. In \cite{1301.0385}, we considered a Einstein-Maxwell-scalar system and
studied its holographic dual QCD model. We obtained a family of analytic black
hole solutions by the potential reconstruction method. By studying the
thermodynamics of the black hole backgrounds, we found a phase transition
between two black holes with different size. We interpreted this black hole to
black hole phase transition as the confinement-deconfinement phase transition
of heavy quarks in the dual holographic QCD model.

On the other hand, the heavy quark potential is an important observable
relevant to confinement. It has been measured in great detail in lattice
simulations \cite{2001} and the results remarkably agree with the Cornell
potential \cite{Cornel}%
\begin{equation}
V\left(  r\right)  =-\dfrac{\kappa}{r}+\sigma_{s}r+C,
\end{equation}
which is dominant by Coulomb potential at short distances and by linear
potential at large distances with the coefficient $\sigma_{s}$ defined as
string tension. In QCD, the heavy quark potential can be read off from the
expectation value of the Wilson loop along a time-like closed path $C$,%
\begin{equation}
\left\langle W\left(  C\right)  \right\rangle \sim e^{-tV\left(  r\right)  }.
\end{equation}
In string/gauge duality, the expectation value of the Wilson loop is given by
\cite{9803002}%
\begin{equation}
\left\langle W\left(  C\right)  \right\rangle =\int DXe^{-S_{NG}},
\end{equation}
where $S_{NG}$ is the string world-sheet action bounded by the loop $C$ at the
boundary of an AdS space. In
\cite{9803135,9803137,0604204,0610135,0611304,0701157,0807.4747,1004.1880,1008.3116,1201.0820,1206.2824,1401.3635}%
, a probe open string in an AdS background was considered. The two ends of the
open string are attached to the boundary of AdS background and behave as a
quark-antiquark pair. Thus the open string could be interpreted as a bound
state, i.e. meson state, in QCD. By studying the dynamics of the open string,
the expectation value of the Wilson loop can be obtained, so as the heavy
quark potential. From the behavior of the heavy quark potential, one is able
to study the process that an open string breaks to two open strings with their
two ends attaching to the AdS boundary and the black hole horizon,
respectively. This string breaking phenomenon describes\ how a meson melts to
a pair of free quark and antiquark in its dual QCD.

\begin{figure}[h]
\begin{center}
\includegraphics[
height=1.727in,
width=5.74in
]{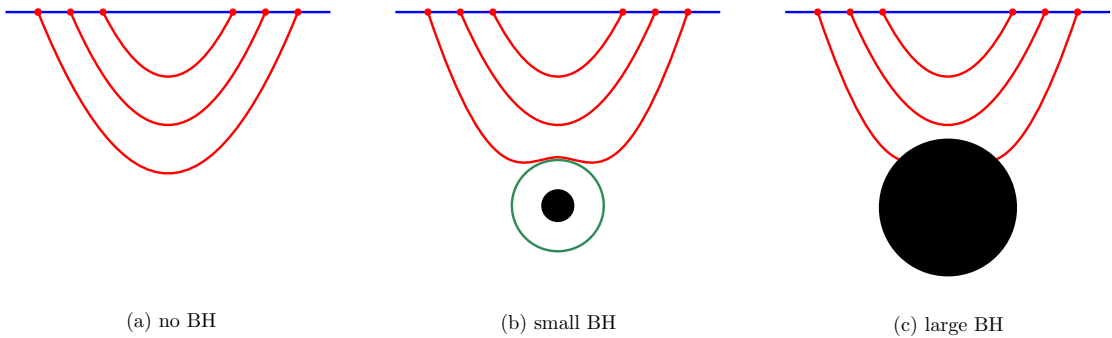}
\end{center}
\caption{ Three configurations for open strings in a black hole. There is no
black hole in case (a), open strings are always connected with their two ends
on the AdS boundary. In the small black hole case (b), open strings can not
exceed a certain distance from the boundary and are still connected with their
two ends on the AdS boundary. In the large black hole case (c), the open
strings with their two ends far away enough will break to two open strings
with their two ends attaching the AdS boundary and the black hole horizon,
respectively.}%
\label{string-black hole}%
\end{figure}

In this work, we put probe open strings in the background obtained in
\cite{1301.0385}. We study the dynamics of the open strings to obtain the
expectation value of the Wilson loop as well as the heavy quark potential. In
\cite{1301.0385}, various black hole phases for different temperatures have
been obtained. In this work, we found three open string configurations for the
various black hole phases as in figure \ref{string-black hole}. According to
AdS/QCD duality, these different open string configurations correspond to the
confinement and deconfinement phases in QCD, respectively. This supports our
preferred interpretation that the black hole to black hole phase transition in
the bulk corresponds to the confinement-deconfinement phase transition of
heavy quarks in the dual holographic QCD in \cite{1301.0385}. Nevertheless, we
found that the phase transition temperatures obtained from the black hole
phases and the open string configurations are not exact the same. In fact, we
will argue that neither the black hole phases nor the string configurations
alone could explain the full phase structure of the confinement-deconfinement
phase transition in QCD. The string configurations tells us that whether the
system is in confinement or deconfinement phase, while the black hole phase
transition determines the location of the phase boundary. By combining the two
effects together in this paper, we find a more natural picture to describe the
phase diagram of the confinement-deconfinement transition for the heavy quarks
in QCD. Furthermore, in the deconfinement phase, we also study the meson
melting process by studying the process of an open string breaking to two open strings.

The paper is organized as follows. In section II, we consider an
Einstein-Maxwell-scalar system. We review how to get the analytic solutions in
\cite{1301.0385} by potential reconstruction method and study the phase
structure in the these backgrounds. In section III, we add probe open strings
in our black hole background to study their various configurations. We
calculate the expectation value of the Wilson loop and study the heavy quark
potential. In section IV, by combining the background phase structure and the
open string breaking effect, we obtain the phase diagram for
confinement-deconfinement transition. We further study the meson melting
process in the deconfinement phase. We conclude our result in section IV.%

\setcounter{equation}{0}
\renewcommand{\theequation}{\arabic{section}.\arabic{equation}}%

\section{Einstein-Maxwell-Scalar System}

In this section, we review the black hole solution and its phase structure
obtained in \cite{1301.0385}.

\subsection{Background}

We consider a 5-dimensional Einstein-Maxwell-scalar system with probe matters.
The action of the system has two parts, the background part and the matter
part,%
\begin{equation}
S=S_{b}+S_{m}.
\end{equation}
In Einstein frame, the background action includes a gravity field $g_{\mu\nu}%
$, a Maxwell field $A_{\mu}$ and a neutral scalar field $\phi$, while the
matter action includes a massless gauge fields ${A}_{\mu}^{V}$, which we will
treat as probe, describing the degrees of freedom of vector mesons on the 4d
boundary,%
\begin{align}
S_{b}  &  =\dfrac{1}{16\pi G_{5}}\int d^{5}x\sqrt{-g}\left[  {R-\frac{f\left(
\phi\right)  }{4}F^{2}}-\dfrac{1}{2}\partial_{\mu}\phi\partial^{\mu}%
\phi-V\left(  \phi\right)  \right]  ,\label{action-b}\\
S_{m}  &  =-\dfrac{1}{16\pi G_{5}}\int d^{5}x\sqrt{-g}{\frac{f\left(
\phi\right)  }{4}}F_{V}^{2}, \label{action-m}%
\end{align}
where ${G}_{5}$ is the coupling constant for the gauge field strength
${F}_{\mu\nu}{=\partial}_{\mu}A_{\nu}-{\partial}_{\nu}A_{\mu}$, $f\left(
\phi\right)  $ is the gauge kinetic function associated to the Maxwell field
$A_{\mu}$ and $V\left(  \phi\right)  $ is the potential of the scalar field
$\phi$.

The equations of motion can be derived from the above action as%
\begin{align}
\nabla^{2}\phi &  =\frac{\partial V}{\partial\phi}+\frac{1}{4}\frac{\partial
f}{\partial\phi}\left(  F^{2}+{F_{V}^{2}}\right)  ,\text{ \ }\nabla_{\mu
}\left[  f(\phi)F^{\mu\nu}\right]  ={{0,}}\text{ \ }\nabla_{\mu}\left[
f(\phi)F_{V}^{\mu\nu}\right]  ={{0,}}\\
R_{\mu\nu}-\frac{1}{2}g_{\mu\nu}R  &  =\frac{f(\phi)}{2}\left(  F_{\mu\rho
}F_{\nu}^{\rho}-\frac{1}{4}g_{\mu\nu}F^{2}\right)  +\frac{1}{2}\left[
\partial_{\mu}\phi\partial_{\nu}\phi-\frac{1}{2}g_{\mu\nu}\left(  \partial
\phi\right)  ^{2}-g_{\mu\nu}V\right]  .
\end{align}
To solve the background of the Einstein-Maxwell-scalar system, we first turn
off the probe gauge field ${A}_{\mu}^{V}$ and consider the ansatz for the
metric, scalar field and Maxwell field as,%
\begin{align}
ds^{2}  &  =\dfrac{e^{2A\left(  z\right)  }}{z^{2}}\left[  -g(z)dt^{2}%
+\frac{dz^{2}}{g(z)}+d\vec{x}^{2}\right]  ,\label{metric}\\
\phi &  =\phi\left(  z\right)  \text{, \ \ }A_{\mu}=A_{t}\left(  z\right)  ,
\label{ansatz}%
\end{align}
which leads to the following equations of motion for the background fields,%
\begin{align}
\phi^{\prime\prime}+\left(  \frac{g^{\prime}}{g}+3A^{\prime}-\dfrac{3}%
{z}\right)  \phi^{\prime}+\left(  \frac{z^{2}e^{-2A}A_{t}^{\prime2}f_{\phi}%
}{2g}-\frac{e^{2A}V_{\phi}}{z^{2}g}\right)   &  =0,\label{eom-phi}\\
A_{t}^{\prime\prime}+\left(  \frac{f^{\prime}}{f}+A^{\prime}-\dfrac{1}%
{z}\right)  A_{t}^{\prime}  &  =0,\label{eom-At}\\
A^{\prime\prime}-A^{\prime2}+\dfrac{2}{z}A^{\prime}+\dfrac{\phi^{\prime2}}{6}
&  =0,\label{eom-A}\\
g^{\prime\prime}+\left(  3A^{\prime}-\dfrac{3}{z}\right)  g^{\prime}%
-e^{-2A}z^{2}fA_{t}^{\prime2}  &  =0,\label{eom-g}\\
A^{\prime\prime}+3A^{\prime2}+\left(  \dfrac{3g^{\prime}}{2g}-\dfrac{6}%
{z}\right)  A^{\prime}-\dfrac{1}{z}\left(  \dfrac{3g^{\prime}}{2g}-\dfrac
{4}{z}\right)  +\dfrac{g^{\prime\prime}}{6g}+\frac{e^{2A}V}{3z^{2}g}  &  =0.
\label{eom-V}%
\end{align}
To solve the above equations of motion, we need to specify the following
boundary and physical conditions:

\begin{enumerate}
\item Near the boundary $z\rightarrow0$, we require the metric in string frame
to be asymptotic to $AdS_{5}$;

\item Near the horizon $z=z_{H}$, we put the regular condition $A_{t}\left(
z_{H}\right)  =g\left(  z_{H}\right)  =0$;

\item The vector meson spectrum should satisfy the linear Regge trajectories
at zero temperature and zero density \cite{0507246}.
\end{enumerate}

With the above conditions, the equations of motion (\ref{eom-phi}-\ref{eom-V})
can be analytically solved as%
\begin{align}
\phi^{\prime}\left(  z\right)   &  =\sqrt{-6\left(  A^{\prime\prime}%
-A^{\prime2}+\dfrac{2}{z}A^{\prime}\right)  },\label{phip-A}\\
A_{t}\left(  z\right)   &  =\mu\dfrac{e^{cz^{2}}-e^{cz_{H}^{2}}}%
{1-e^{cz_{H}^{2}}},\label{At-A}\\
g\left(  z\right)   &  =1+\dfrac{1}{\int_{0}^{z_{H}}y^{3}e^{-3A}dy}\left[
\dfrac{2c\mu^{2}}{\left(  1-e^{cz_{H}^{2}}\right)  ^{2}}\left\vert
\begin{array}
[c]{cc}%
\int_{0}^{z_{H}}y^{3}e^{-3A}dy & \int_{0}^{z_{H}}y^{3}e^{-3A}e^{cy^{2}}dy\\
\int_{z_{H}}^{z}y^{3}e^{-3A}dy & \int_{z_{H}}^{z}y^{3}e^{-3A}e^{cy^{2}}dy
\end{array}
\right\vert -\int_{0}^{z}y^{3}e^{-3A}dy\right]  ,\\
V\left(  z\right)   &  =-3z^{2}ge^{-2A}\left[  A^{\prime\prime}+3A^{\prime
2}+\left(  \dfrac{3g^{\prime}}{2g}-\dfrac{6}{z}\right)  A^{\prime}-\dfrac
{1}{z}\left(  \dfrac{3g^{\prime}}{2g}-\dfrac{4}{z}\right)  +\dfrac
{g^{\prime\prime}}{6g}\right]  , \label{V-A}%
\end{align}
where $\mu\equiv A_{t}\left(  0\right)  $ is defined as chemical potential.

The solution (\ref{phip-A}-\ref{V-A}) depends on the warped factor $A\left(
z\right)  $. The choice of $A\left(  z\right)  $ is arbitrary provided it
satisfies the boundary conditions. To be concrete, we fix the warped factor
$A\left(  z\right)  $ in our solution in a simple form as%
\begin{equation}
A\left(  z\right)  =-\dfrac{c}{3}z^{2}-bz^{4}, \label{A}%
\end{equation}
where the parameters $b$ and $c$ will be determined later.

\subsection{Phase Structure of the Background}

With the background (\ref{metric}), one can calculate the Hawking-Bekenstein
entropy%
\begin{equation}
s=\dfrac{e^{3A\left(  z_{H}\right)  }}{4z_{H}^{3}}, \label{entropy}%
\end{equation}
and the Hawking temperature%
\begin{equation}
T=\dfrac{z_{H}^{3}e^{-3A\left(  z_{H}\right)  }}{4\pi\int_{0}^{z_{H}}%
y^{3}e^{-3A}dy}\left[  1-\dfrac{2c\mu^{2}\left(  e^{cz_{H}^{2}}\int_{0}%
^{z_{H}}y^{3}e^{-3A}dy-\int_{0}^{z_{H}}y^{3}e^{-3A}e^{cy^{2}}dy\right)
}{\left(  1-e^{cz_{H}^{2}}\right)  ^{2}}\right]  .
\end{equation}

\begin{figure}[h]
\begin{center}
\includegraphics[
height=2in,
width=3in
]{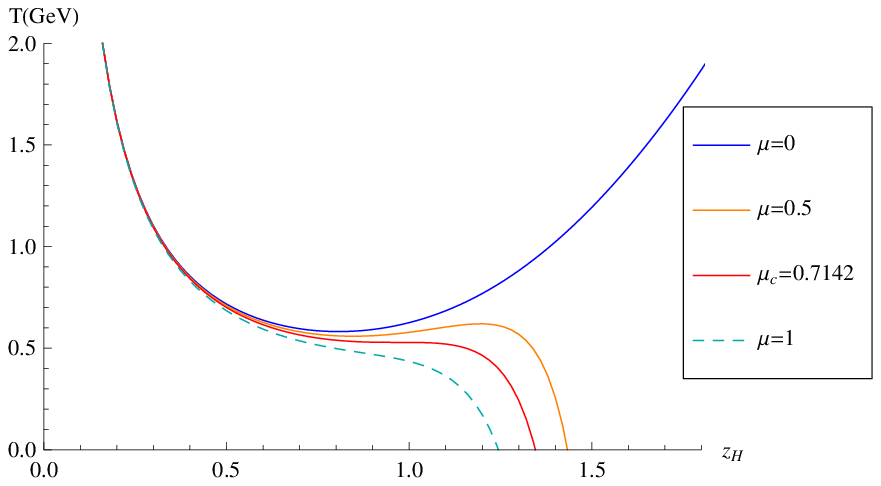}\hspace*{0.5cm} \includegraphics[
height=2in,
width=2.9in
]{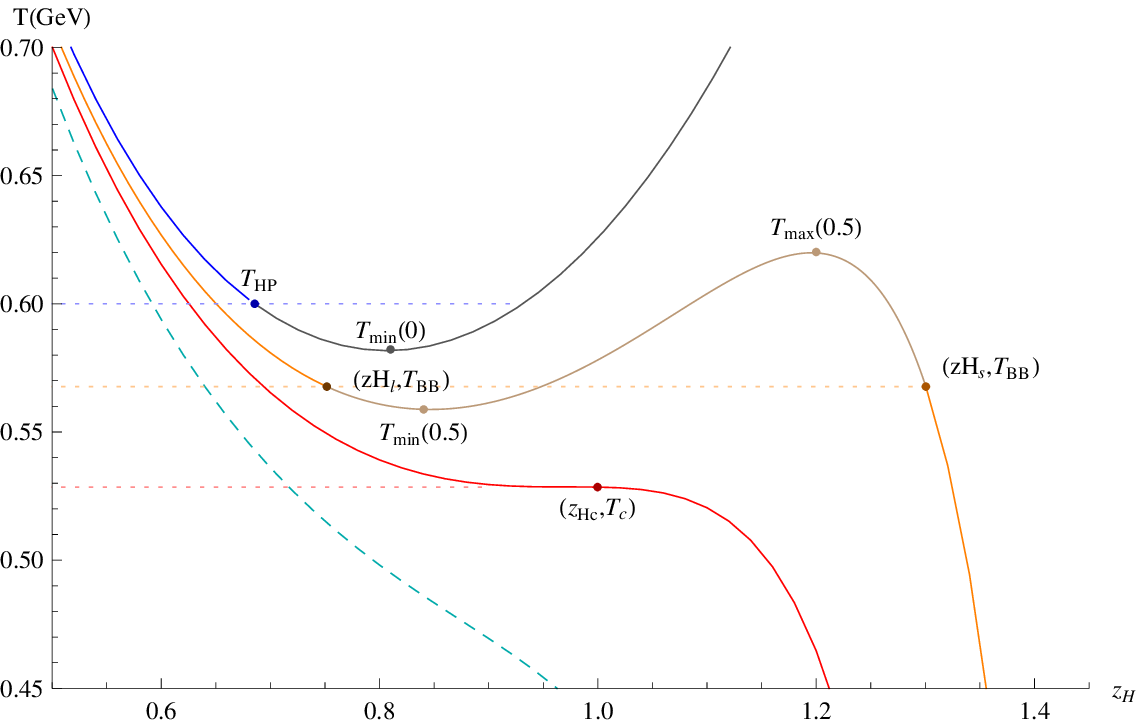}\vskip -0.05cm \hskip 0.15 cm \textbf{( a ) } \hskip 7.5 cm \textbf{(
b )}
\end{center}
\caption{The temperature v.s. horizon at different chemical potentials
$\mu=0,0.5,0.714,1GeV$. We enlarge a rectangle region in (a) into (b) to see
the detailed structure. For $\mu>\mu_{c}$, the temperature decreases
monotonously to zero; while for $\mu<\mu_{c}$, the temperature has a local
minimum. At $\mu_{c}\simeq0.714GeV$, the local minimum reduces to a inflection
point.}%
\label{temperature}%
\end{figure}

\begin{figure}[h]
\begin{center}
\includegraphics[
height=1.7in,
width=3.74in
]{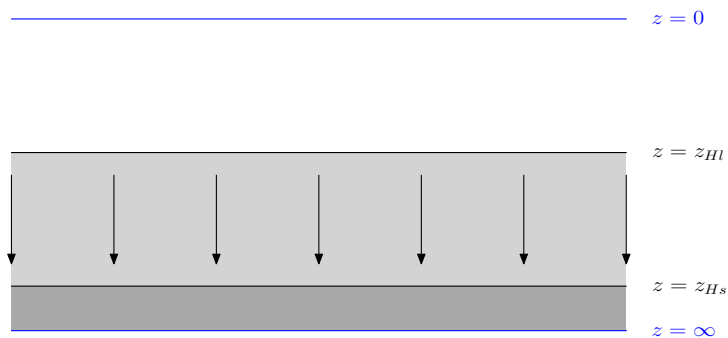}
\end{center}
\caption{Phase transition from a large black hole with the horizon $z=z_{Hl}$
collapses to a small black hole with the horizon $z=z_{Hs}$ at the transition
temperature $T=T_{BB}$.}%
\label{horizon}%
\end{figure}

The temperature $T$ v.s. horizon $z_{H}$ at different chemical potentials is
plotted in figure \ref{temperature}. At $\mu=0$, the temperature has a global
minimum $T_{\min}\left(  0\right)  $ at $z_{H}=z_{\min}\left(  0\right)  $.
The black hole solution is only thermodynamically stable for $z_{H}<z_{\min
}\left(  0\right)  $ and is unstable for $z_{H}>z_{\min}\left(  0\right)  $.
Below the temperature $T_{\min}\left(  0\right)  $, there is no black hole
solution and we expect a Hawking-Page phase transition happens at a
temperature $T_{HP}\left(  0\right)  \gtrsim T_{\min}\left(  0\right)  $ where
the black hole dissolves to a thermal gas background. For $0<\mu<\mu_{c}$, the
temperature has a local minimum/maximum $T_{\min}\left(  \mu\right)  /T_{\max
}\left(  \mu\right)  $\ at $z_{H}=z_{\min}\left(  \mu\right)  /z_{\max}\left(
\mu\right)  $ and decreases to zero at a finite size of horizon. The black
holes between $z_{\min}\left(  \mu\right)  $ and $z_{\max}\left(  \mu\right)
$ are thermodynamically unstable. There are two sections that are stable with
$z_{H}<z_{\min}\left(  \mu\right)  $ and $z_{H}>z_{\max}\left(  \mu\right)  $.
We expect a similar Hawking-Page phase transition happens at a temperature
$T_{HP}\left(  \mu\right)  \gtrsim T_{\min}\left(  \mu\right)  $.
Nevertheless, since the thermodynamically stable black hole solutions exist
even when the temperature below $T_{\min}\left(  \mu\right)  $ for the section
$z_{H}>z_{\max}\left(  \mu\right)  $, we also expect a black hole to black
hole phase transition happening at a temperature $T_{BB}\left(  \mu\right)  $
between $T_{\min}\left(  \mu\right)  $ and $T_{\max}\left(  \mu\right)  $,
where a large black hole with the horizon $z=z_{Hl}\left(  \mu\right)  $
collapses to a small black hole with the horizon $z=z_{Hs}\left(  \mu\right)
$ as showed in figure \ref{horizon}. Finally, for $\mu>\mu_{c}$, the
temperature monotonously decreases to zero and there is no black hole to black
hole phase transition anymore\footnote{There could still be a Hawking-Page
phase transition at some temperature for the case of $\mu>\mu_{c}$, but we
will show later that the black hole solution is always thermodynamically
favored in this case.}.

\begin{figure}[h]
\begin{center}
\includegraphics[
height=2in,
width=3in
]{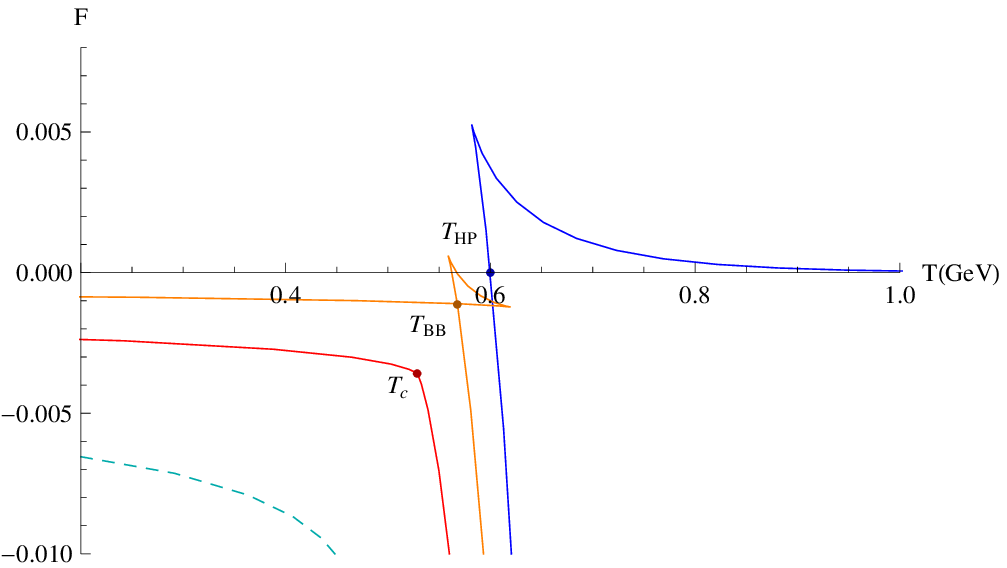}\hspace*{0.5cm} \includegraphics[
height=2in,
width=2.9in
]{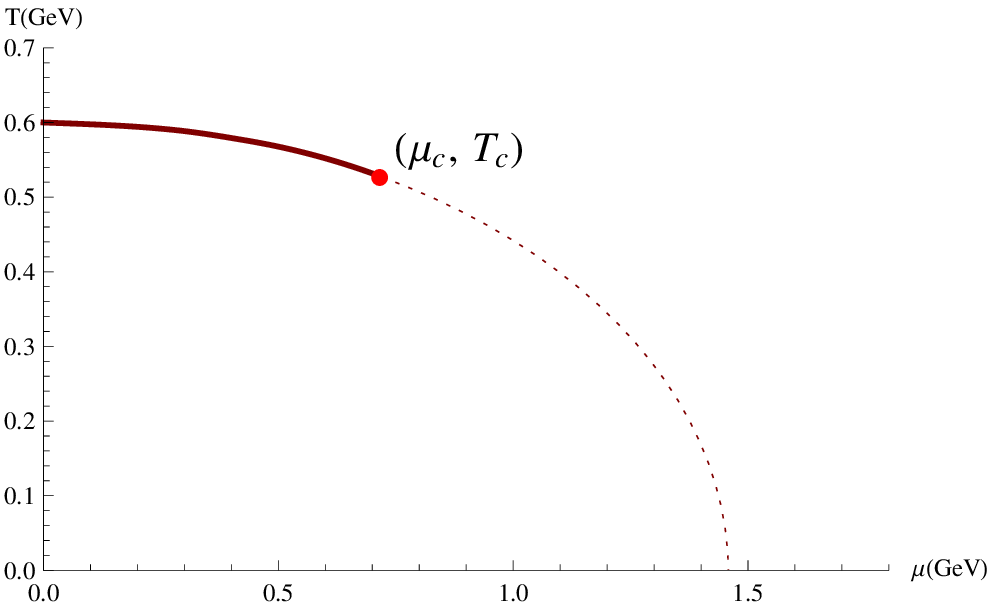}\vskip -0.05cm \hskip 0.15 cm \textbf{( a ) } \hskip 7.5 cm
\textbf{( b )}
\end{center}
\caption{(a) The free energy v.s. temperature at chemical potentials
$\mu=0,0.5,0.714,1 GeV$. At $\mu=0$, the free energy intersect with the
$x$-axis at $T=T_{HP}$ where the black hole dissolves to thermal gas by
Hawking-Page phase transition. For $0<\mu<\mu_{c}\simeq0.714GeV$, the
temperature reaches its maximum value where the free energy turns back to
intersect with itself at $T=T_{BB}$ where the black hole to black hole
transition happens. For $\mu>\mu_{c}$, the swallow-tailed shape disappears and
there is no phase transition in the background. (b) The phase diagram in $T$
and $\mu$ plane. At small $\mu$, the system undergoes a first order phase
transition at finite $T$. The first order phase transition stops at the
critical point $(\mu_{c},T_{c})=(0.714GeV,0.528GeV)$, where the phase
transition becomes second order. For $\mu>\mu_{c}$, the system weaken to a
sharp but smooth crossover \cite{1301.0385}.}%
\label{phase diagram}%
\end{figure}

To determine the phase transition temperatures $T_{HP}\left(  \mu\right)  $
and $T_{BB}\left(  \mu\right)  $, we compute the free energy from the first
law of thermodynamics in grand canonical ensemble%
\begin{equation}
F=-\int sdT. \label{int F}%
\end{equation}
We plot the free energy v.s. temperature in (a) of figure \ref{phase diagram}.

At $\mu=0$, the free energy intersect the $x$-axis at $T=T_{HP}\left(
0\right)  $ where the Hawking-Page phase transition happens. The black hole
dissolves to thermal gas which is thermodynamically stable for $T<T_{HP}%
\left(  0\right)  $. We fix the parameter $b\simeq0.273GeV^{4}$ in Eq.
(\ref{A}) by fitting the Hawking-Page phase transition temperature
$T_{HP}\left(  0\right)  $ with the lattice QCD simulation of $T_{HP}%
\simeq0.6GeV$ in \cite{1111.4953}.

For $0<\mu<\mu_{c}$, the free energy behaves as the expected swallow-tailed
shape. The temperature reaches its maximum where the free energy turns back
and intersects with itself at $T=T_{BB}\left(  \mu\right)  $ where the large
black hole transits to the small black hole. Since the free energies of the
stable black holes are always less than that of the thermal gas ($F_{gas}%
\equiv0$), the thermodynamic system will always favor the small black hole
background other than the thermal gas background. When we increase the
chemical potential $\mu$ from zero to $\mu_{c}$, the loop of the
swallow-tailed shape shrinks to disappear at $\mu=\mu_{c}$. For $\mu>\mu_{c}$,
the curve of the free energy increases smoothly from higher temperature to
lower temperature.

The phase diagram of the background is plotted in (b) of figure
\ref{phase diagram}. At $\mu=0$, the system undergoes a black hole to thermal
gas phase transition at $T=T_{HP}\left(  0\right)  $. For $0<\mu<\mu_{c}$, the
system undergoes a large black hole to small black hole phase transition at
$T_{BB}\left(  \mu\right)  $. The phase transition temperature $T_{BB}\left(
\mu\right)  $ approaches to $T_{HP}$ at $\mu\rightarrow0$ that makes the phase
diagram continuous at $\mu=0$. The phase transition stops at $\mu=\mu_{c}$ and
reduces to a crossover for $\mu>\mu_{c}$.

The phase diagram we obtained here in figure \ref{phase diagram} is different
from the conventional QCD phase diagram, in which crossover happens for small
chemical potential and phase transition happens for large chemical potential.
In \cite{1301.0385}, by comparing with the phase structure in lattice QCD
simulation, the authors argued that this 'reversed' phase diagram should be
interpreted as confinement-deconfinement phase transition of heavy quarks in
QCD. In this paper, we consider the same background as in \cite{1301.0385} to
study pure gluon QCD with one additional heavy flavour, and not light quarks.
Since our model describes heavy quarks system in QCD, the flavour field
${A}_{\mu}^{V}$ in the matter action \ref{action-m} should be associated to
the mesons make up of heavy quarks, i.e. quarkonium states. By fitting the
lowest two spectrum of quarkonium states $m_{J/\psi}=3.096GeV$ and
$m_{\psi^{\prime}}=3.685GeV$, we can fix $c\simeq1.16GeV^{2}$ in Eq. (\ref{A}).

Nevertheless, there left a problem that, in the gravity side, it is commonly
believed that the confinement-deconfinement phase transition in the field
theory side is dual to the Hawking-Page phase transition. Hawking-Page phase
transition is the transition between black hole and thermal gas backgrounds.
However, in our gravity background, the phase transition is between two black
holes for a non-zero chemical potential. Thus it is not consistent to consider
a black hole to black hole phase transition in the gravity side to be dual to
the confinement-deconfinement phase transition in QCD. In the following of
this paper, by adding open strings in the background, we will study this issue
more carefully to gain a more reasonable physical picture.%

\setcounter{equation}{0}
\renewcommand{\theequation}{\arabic{section}.\arabic{equation}}%

\section{Open Strings in the Background}

\begin{figure}[h]
\begin{center}
\includegraphics[
height=2.3in,
width=2.5in
]{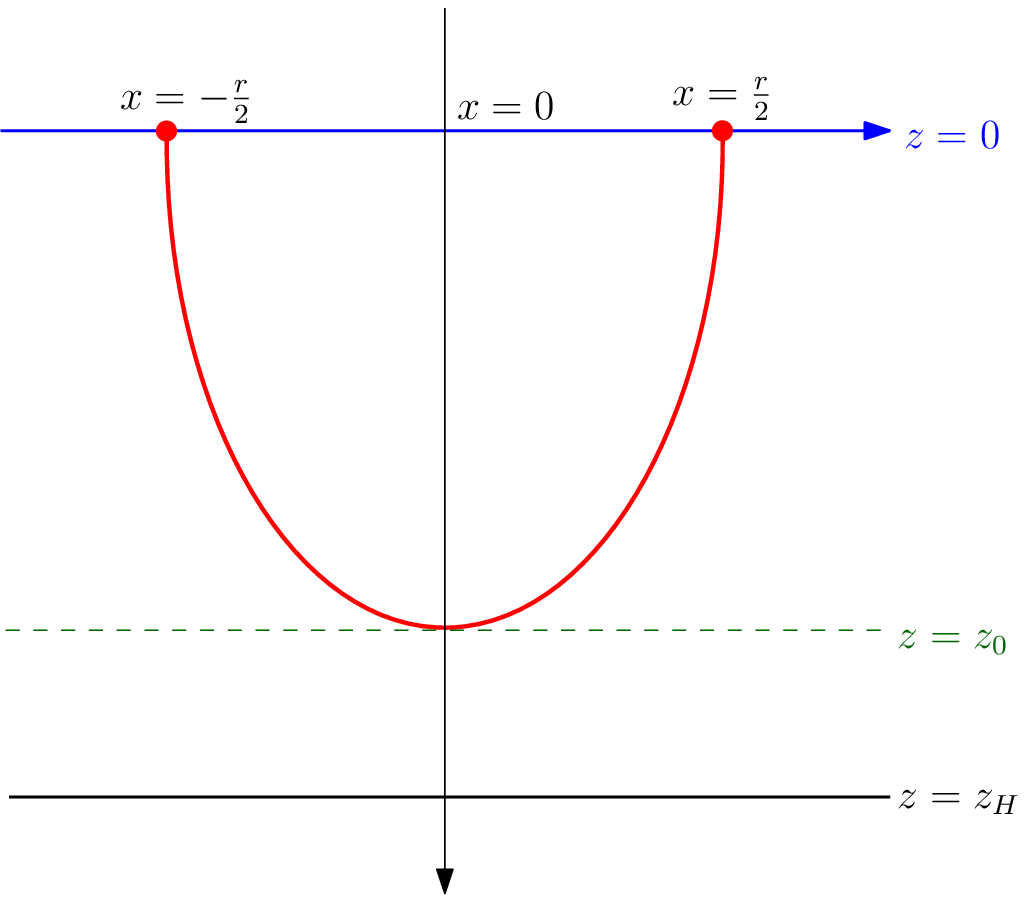}\hspace*{0.7cm} \includegraphics[
height=2.3in,
width=2.5in
]{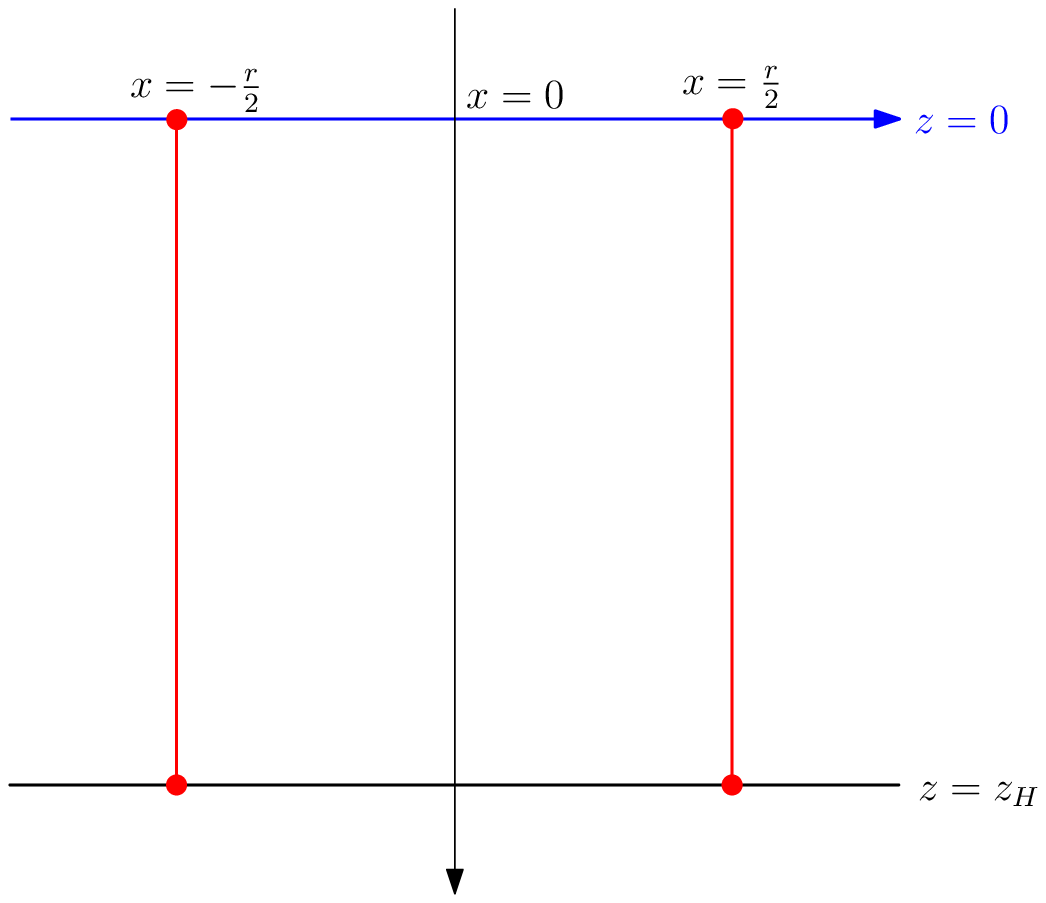}\vskip -0.05cm \hskip 0.12 cm \textbf{( a ) } \hskip 6 cm
\textbf{( b )}
\end{center}
\caption{Two open string configurations. In (a), a U-shape open string
connects its two ends at the boundary at $z=0$ and reaches its maximum depth
at $z=z_{0}$. In (b), two straight open strings with their ends connecting the
boundary at $z=0$ and the horizon at $z=z_{H}$, respectively.}%
\label{string}%
\end{figure}

In this paper, we consider an open string in the above background with its two
ends on the boundary of the space-time at $z=0$. There are two configurations
for an open string in the black hole background. One is the U-shape
configuration with the open string reaching its maximum depth at $z=z_{0}$;
the other is the straight configuration with the straight open string having
its two ends attached to the boundary and the horizon at $z=z_{H}$,
respectively. The two configurations are showed in figure \ref{string}. Since
the dual holographic QCD lives on the boundary, it is natural for us to
interpret the two ends of the open string as a quark-antiquark pair. The
U-shape configuration corresponds to the quark-antiquark pair being connected
by a string and can be identified as a meson state. While the straight
configuration corresponds to a free quark or antiquark.

The Nambu-Goto action of an open string is%
\begin{equation}
S_{NG}=\int d^{2}\xi\sqrt{-G},
\end{equation}
where the induced metric%
\begin{equation}
G_{ab}=g_{\mu\nu}\partial_{a}X^{\mu}\partial_{b}X^{\nu},
\end{equation}
on the 2-dimensional world-sheet that the string sweeps out as it moves with
coordinates $(\xi^{0},\xi^{1})$ is the pullback of 5-d target space-time
metric $g_{\mu\nu}$,%
\begin{equation}
ds^{2}=\frac{e^{2A(z)}}{z^{2}}\left(  g(z)dt^{2}+d\vec{x}^{2}+\frac{1}%
{g(z)}dz^{2}\right)  ,
\end{equation}
where, to study the thermal properties of the system, we consider the
Euclidean metric and identify the periodic of the time with the inverse of
temperature as $\beta=1/T$.

\subsection{Wilson Loop}

\begin{figure}[h]
\begin{center}
\includegraphics[
height=1.4905in,
width=4.4823in
]{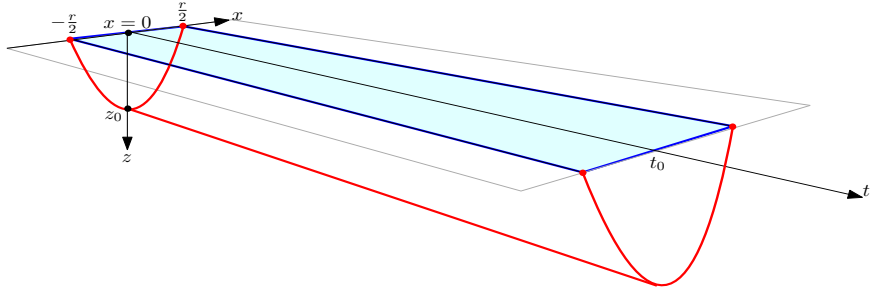}
\end{center}
\caption{Wilson loop as the boundary of string world-sheet.}%
\label{Wilson loop}%
\end{figure}

We consider a $r\times t_{0}$ rectangular Wilson loop $C$ along the directions
$\left(  t,x\right)  $\ on the boundary of the AdS space attached by a pair of
the quark and antiquark separated by $r$. The quark and antiquark located at
$\left(  z=0,x=\pm r/2\right)  $ are connected by an open string, which
reaches its maximum at $\left(  z=z_{0},x=0\right)  $ as in figure
\ref{Wilson loop}.

It is known that taking the limit $t_{0}\rightarrow\beta=1/T$ allows one to
read off the energy of such a pair from the expectation value of the Wilson
loop,%
\begin{equation}
\left\langle W\left(  C\right)  \right\rangle \sim e^{-V\left(  r,T\right)
/T},
\end{equation}
where $V\left(  r,T\right)  $ is the heavy quark-antiquark potential
\cite{0604204,1201.0820}.

In string/gauge duality, the expectation value of the Wilson loop is given by%
\begin{equation}
\left\langle W\left(  C\right)  \right\rangle =\int DXe^{-S_{NG}}\simeq
e^{-S_{on-shell}}, \label{Wilson}%
\end{equation}
where $S_{NG}$ is the string world-sheet action bounded by a curve $C$ at the
boundary of AdS space and $S_{on-shell}$ is the on-shell string action, which
is proportional to the area of the string world-sheet bounded by the Wilson
loop $C$.

Comparing with Eq. (\ref{Wilson}), the free energy of the meson is defined as%
\begin{equation}
V\left(  r,T\right)  =TS_{on-shell}\left(  r,T\right)  .
\end{equation}

\subsection{Configurations of Open Strings}

\begin{figure}[h]
\begin{center}
\includegraphics[
height=2.5in,
width=4in
]{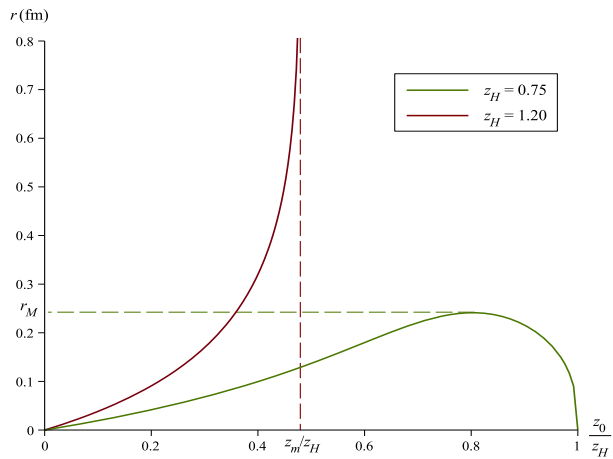}
\end{center}
\caption{Separate distance $r$ between quark and antiquark v.s. $z_{0}$ at
$\mu=0.5GeV$.}%
\label{r-z0}%
\end{figure}

The string world-sheet action is defined by the Nambu-Goto action,%
\begin{equation}
S=\int d^{2}\xi\mathcal{L}=\int d^{2}\xi\sqrt{\det G},
\end{equation}
where $G_{ab}=\partial_{a}X^{\mu}\partial_{b}X_{\mu}$\ is the induced metric
on string world-sheet. For the meson configuration, by choosing static gauge:
$\xi^{0}=t,$ $\xi^{1}=x$, the induced metric in string frame becomes%
\begin{equation}
ds^{2}=G_{ab}d\xi^{a}d\xi^{b}=\frac{e^{2A\left(  z\right)  }}{z^{2}}g\left(
z\right)  dt^{2}+\frac{e^{2A\left(  z\right)  }}{z^{2}}\left(  1+\dfrac
{z^{\prime2}}{g\left(  z\right)  }\right)  dx^{2},
\end{equation}
where the prime denotes a derivative with respect to $x$. The Lagrangian and
Hamiltonian can be calculated as%
\begin{align}
\mathcal{L}  &  =\sqrt{\det G}=\frac{e^{2A\left(  z\right)  }}{z^{2}}%
\sqrt{g\left(  z\right)  +z^{\prime2}},\\
\mathcal{H}  &  =\left(  \frac{\partial\mathcal{L}}{\partial z^{\prime}%
}\right)  z^{\prime}-\mathcal{L}=-\frac{e^{2A\left(  z\right)  }g\left(
z\right)  }{z^{2}\sqrt{g\left(  z\right)  +z^{\prime2}}}. \label{H}%
\end{align}
With boundary conditions%
\begin{equation}
z\left(  x=\pm\frac{r}{2}\right)  =0\text{, }z(x=0)=z_{0}\text{, }z^{\prime
}(x=0)=0,
\end{equation}
we obtain the conserved energy%
\begin{equation}
\mathcal{H}(x=0)=-\frac{e^{2A\left(  z_{0}\right)  }}{z_{0}^{2}}\sqrt{g\left(
z_{0}\right)  }.
\end{equation}
We can solve $z^{\prime}$\ from Eq. (\ref{H}),%
\begin{equation}
z^{\prime}=\sqrt{g\left(  \dfrac{\sigma^{2}\left(  z\right)  }{\sigma
^{2}\left(  z_{0}\right)  }-1\right)  },
\end{equation}
where%
\begin{equation}
\sigma\left(  z\right)  =\frac{e^{2A\left(  z\right)  }\sqrt{g\left(
z\right)  }}{z^{2}}.
\end{equation}
The distance $r$ between the quark-antiquark pair can be calculated as,%
\begin{equation}
r=\int_{-\frac{r}{2}}^{\frac{r}{2}}dx=2\int_{0}^{z_{0}}dz\frac{1}{z^{\prime}%
}=2\int_{0}^{z_{0}}dz\left[  g\left(  z\right)  \left(  \dfrac{\sigma
^{2}\left(  z\right)  }{\sigma^{2}\left(  z_{0}\right)  }-1\right)  \right]
^{-\frac{1}{2}},
\end{equation}
where $z_{0}$ is the maximum depth that the string can reach. The dependence
of the distance $r$ on $z_{0}$ at two different horizons are plotted in figure
\ref{r-z0}.

\begin{figure}[h]
\begin{center}
\includegraphics[
height=1.7in,
width=3in
]{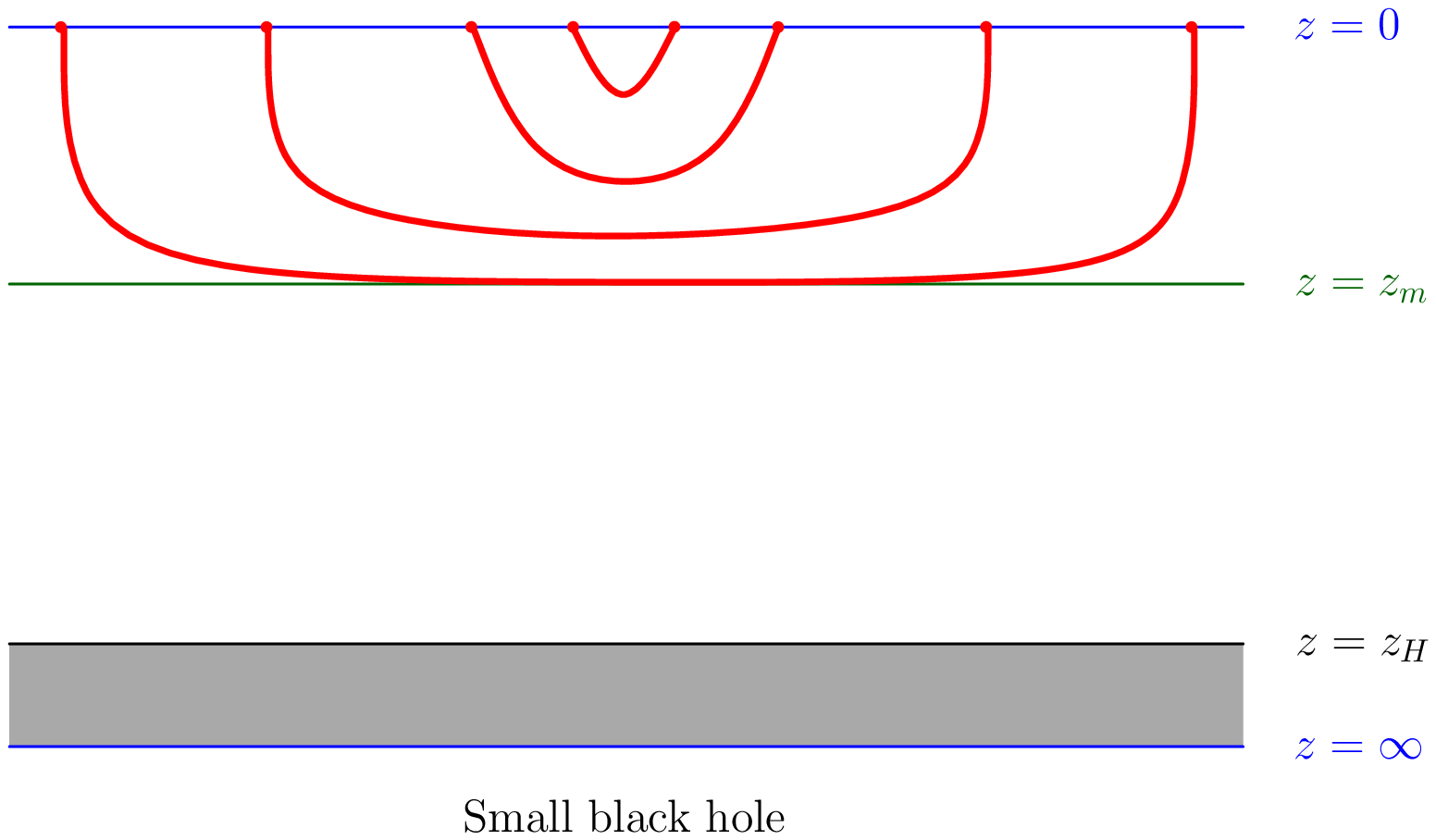}\hspace*{0.5cm} \includegraphics[
height=1.7in,
width=3in
]{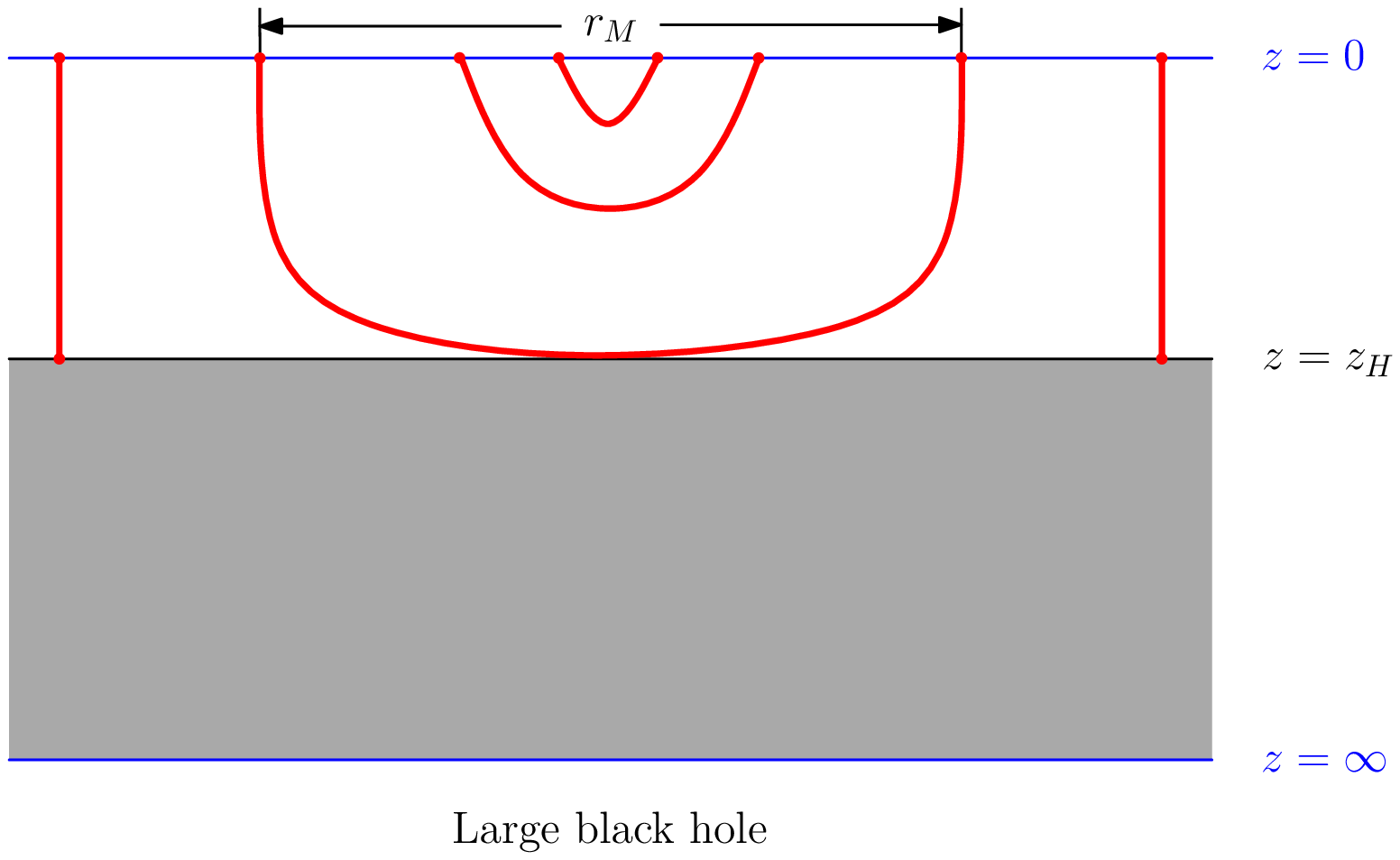}\vskip -0.05cm \hskip 0.13 cm \textbf{( a ) } \hskip 7 cm
\textbf{( b )}
\end{center}
\caption{(a) For a small black hole, open strings can not exceed the dynamical
wall at $z=z_{m}$ and are always in the U-shape. (b) For a large black hole,
an open string will break to two straight strings if the distance between its
two ends is larger than $z_{M}$.}%
\label{conf-deconf}%
\end{figure}

We see that for a small black hole (large $z_{H}$), there exist a dynamical
wall at $z_{m}<z_{H}$ where $r^{\prime}\left(  z_{m}\right)  \rightarrow
\infty$. The open string can not go beyond this dynamical wall, i.e.
$z_{0}\leq z_{m}$, even when the distance $r$ between the quark-antiquark pair
goes to infinity as showed in (a) of figure \ref{conf-deconf}. While for a
large black hole (small $z_{H}$), the open string can reach arbitrary close to
the horizon, but there is a maximum value for the distance at $r=r_{M}$. If
the distance between the quark and antiquark is larger than $r_{M}$,\ there is
no stable U-shape solution for open strings so that the open string of U-shape
will break to two straight open strings connecting the boundary at $z=0$ and
the horizon at $z=z_{H}$ as showed in (b) of figure \ref{conf-deconf}.

\begin{figure}[h]
\begin{center}
\includegraphics[
height=1.7409in,
width=3.755in
]{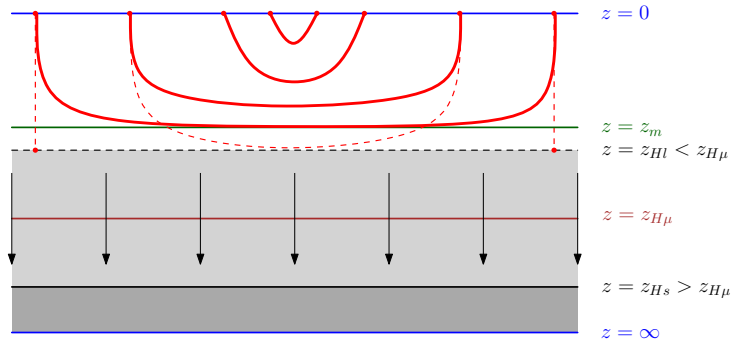}
\end{center}
\caption{When a large black hole with horizon $z_{Hl}<z_{H\mu}$ shrinks to a
small black hole with horizon $z_{Hs}>z_{H\mu}$, the dynamical wall at
$z=z_{m}$ appears when $z_{H}=z_{H\mu}$.}%
\label{phase-cross}%
\end{figure}

In summary, for a small black hole, open strings are always in the U-shape;
while for a large black hole, an open string is in the U-shape for short
separate distance $r<r_{M}$ and is in the straight shape for long separate
distance $r>r_{M}$. Thus when a large black hole shrinks to a small one, we
expect that a dynamical wall will appear when the black hole horizon equal to
a critical value $z_{H\mu}$ for each chemical potential $\mu$, showed in
figure \ref{phase-cross}.

\begin{figure}[h]
\begin{center}
\includegraphics[
height=2.5in,
width=3.7in
]{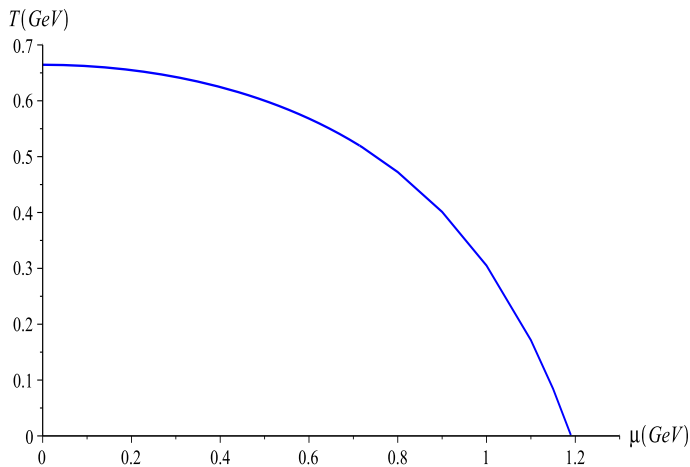}
\end{center}
\caption{ The temperature $T_{\mu}$ corresponding to $z_{H\mu}$, at which the
dynamical wall appears/disaapears, for each chemical potential $\mu$.}%
\label{Tmu}%
\end{figure}

Since each horizon is associated to a temperatures for black holes, we define
the transformation temperature $T_{\mu}$ corresponding to the critical black
hole horizon $z_{H\mu}$, at which the dynamical wall appears/disappears, for
each chemical potential $\mu$. The dependence of $T_{\mu}$ on $\mu$ is plotted
in figure \ref{Tmu}.

When there is dynamical wall, open strings are always in the U-shape. This
means that quarks and antiquarks are always connected by an open string to
form a bound state, i.e. the meson state in QCD. It is natural to interpret
this case as the confinement phase in the dual holographic QCD. On the other
hand, when the dynamical wall disappears, an U-shape open string could break
up to two strings if the distance between its two ends is large enough. This
means that a meson state could break to a pair of free quark and antiquark in
QCD. We interpret this case as the deconfinement phase in the dual holographic
QCD. Therefore, the transformation temperature $T_{\mu}$ is associated to the
transformation between the confinement and the deconfinement phases in the
dual holographic QCD\footnote{This transformation between confinement and
deconfinement phases is not necessary a phase transition, it could be a smooth
crossover as we will show later.}.

\subsection{Heavy Quark Potential}

\begin{figure}[h]
\begin{center}
\includegraphics[
height=2in,
width=3in
]{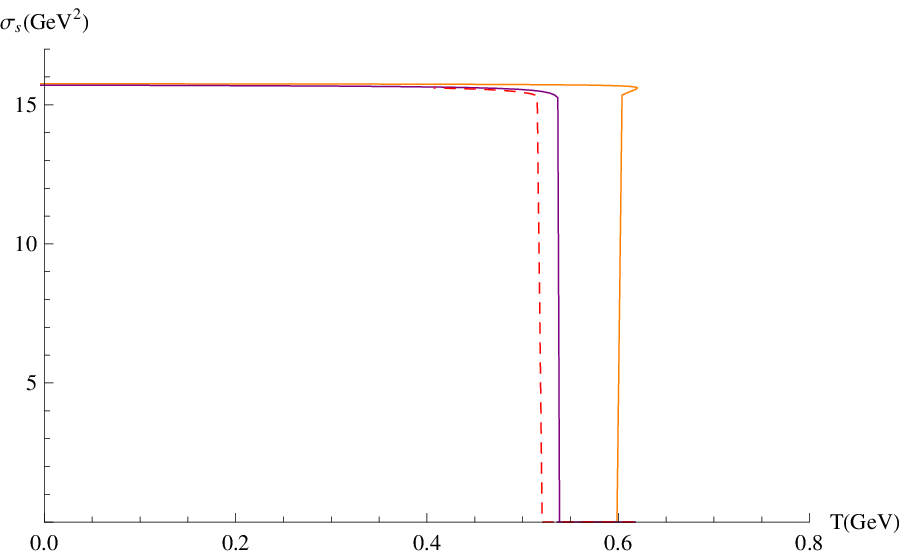}\hspace*{0.5cm} \includegraphics[
height=2in,
width=3in
]{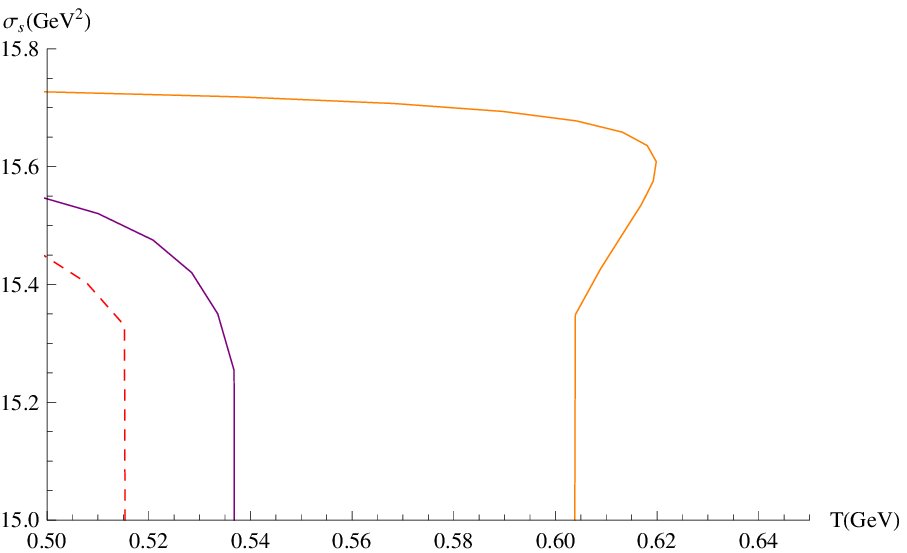}\vskip -0.05cm \hskip 0.15 cm \textbf{( a ) } \hskip 7.5 cm
\textbf{( b )}
\end{center}
\caption{String tension v.s. temperature at $\mu=0.5,0.678,0.714GeV$. (a) The
string tension decreases with temperature growing in the confinement phase,
and suddenly drops to zero at $T=T_{\mu}$. The region closes to the transition
temperatures is enlarged in (b)}%
\label{string tension}%
\end{figure}

When the black hole horizon $z_{H}>z_{H\mu}$, there exists a dynamical wall at
$z=z_{m}<z_{H}$. Open strings are always in the U-shape and the heavy quark
potential can be calculated as%
\begin{equation}
V=TS_{on-shell}=\int_{-\frac{r}{2}}^{\frac{r}{2}}dx\mathcal{L}=2\int%
_{0}^{z_{0}}dz\frac{\sigma\left(  z\right)  }{\sqrt{g\left(  z\right)  }%
}\left[  1-\dfrac{\sigma^{2}\left(  z_{0}\right)  }{\sigma^{2}\left(
z\right)  }\right]  ^{-\frac{1}{2}}. \label{quark potential}%
\end{equation}
In short separate distance limit $r\rightarrow0$, i.e. $z_{0}\rightarrow0$, we
expand the distance and the heavy quark potential at $z_{0}=0$,%
\begin{align}
r  &  =2\int_{0}^{z_{0}}dz\left[  g\left(  z\right)  \left(  \dfrac{\sigma
^{2}\left(  z\right)  }{\sigma^{2}\left(  z_{0}\right)  }-1\right)  \right]
^{-\frac{1}{2}}=r_{1}z_{0}+O(z_{0}^{2}),\\
V  &  =2\int_{0}^{z_{0}}dz\frac{\sigma\left(  z\right)  }{\sqrt{g\left(
z\right)  }}\left[  1-\dfrac{\sigma^{2}\left(  z_{0}\right)  }{\sigma
^{2}\left(  z\right)  }\right]  ^{-\frac{1}{2}}=\frac{V_{-1}}{z_{0}}+O(1),
\end{align}
where%
\begin{align}
r_{1}  &  =2\int_{0}^{1}dv\left(  \frac{1}{v^{4}}-1\right)  ^{-\frac{1}{2}%
}=\frac{1}{2}B\left(  \frac{3}{4},\frac{1}{2}\right)  ,\\
V_{-1}  &  =2\int_{0}^{1}\frac{dv}{v^{2}}\left(  1-v^{4}\right)  ^{-\frac
{1}{2}}=\frac{1}{2}B\left(  -\frac{1}{4},\frac{1}{2}\right)  .
\end{align}
This gives the expected Coulomb potential at short separate distance,%
\begin{equation}
V=-\frac{\kappa}{r},
\end{equation}
where%
\begin{equation}
\kappa=-\dfrac{1}{4}B\left(  \frac{3}{4},\frac{1}{2}\right)  B\left(
-\frac{1}{4},\frac{1}{2}\right)  \simeq1.44.
\end{equation}
In long separate distance $r\rightarrow\infty$, i.e. $z_{0}\rightarrow z_{m}$,
we make a coordinate transformation $z=z_{0}-z_{0}w^{2}$. The distance $r$ and
the heavy quark potential $V$\ become%
\begin{align}
r  &  =2\int_{0}^{1}f_{r}\left(  w\right)  dw,\label{r}\\
V  &  =2\int_{0}^{1}f_{V}\left(  w\right)  dw, \label{FR}%
\end{align}
where%
\begin{align}
f_{r}\left(  w\right)   &  =2z_{0}w\left[  g\left(  z_{0}-z_{0}w^{2}\right)
\left(  \dfrac{\sigma^{2}\left(  z_{0}-z_{0}w^{2}\right)  }{\sigma^{2}\left(
z_{0}\right)  }-1\right)  \right]  ^{-\frac{1}{2}},\\
f_{V}\left(  w\right)   &  =2z_{0}w\frac{\sigma\left(  z_{0}-z_{0}%
w^{2}\right)  }{\sqrt{g\left(  z_{0}-z_{0}w^{2}\right)  }}\left[
1-\dfrac{\sigma^{2}\left(  z_{0}\right)  }{\sigma^{2}\left(  z_{0}-z_{0}%
w^{2}\right)  }\right]  ^{-\frac{1}{2}}.
\end{align}
We learn from figure \ref{r-z0} that the distance $r$ is divergent at
$z_{0}=z_{m}$, and the same happens for the heavy quark potential.\ By
carefully analysis, we find that this divergence is due to the integrands
$f_{r}\left(  w\right)  $ and $f_{r}\left(  w\right)  $ are divergent near the
lower limit $w=0$, i.e. $z=z_{0}\rightarrow z_{m}$. To study the behaviors of
the distance and the heavy quark potential at $z_{0}=z_{m}$, we expand
$f_{r}\left(  w\right)  $\ and $f_{V}\left(  w\right)  $ at $w=0$,%
\begin{align}
f_{r}\left(  w\right)   &  =2z_{0}\left[  -2z_{0}g\left(  z_{0}\right)
\frac{\sigma^{\prime}\left(  z_{0}\right)  }{\sigma\left(  z_{0}\right)
}\right]  ^{-\frac{1}{2}}+O\left(  w\right)  ,\\
f_{V}\left(  w\right)   &  =2z_{0}\sigma\left(  z_{0}\right)  \left[
-2z_{0}g\left(  z_{0}\right)  \frac{\sigma^{\prime}\left(  z_{0}\right)
}{\sigma\left(  z_{0}\right)  }\right]  ^{-\frac{1}{2}}+O\left(  w\right)  .
\end{align}
The integrals (\ref{r}) and (\ref{FR}) can be approximated by only consider
the leading terms of $f_{r}\left(  w\right)  $ and $f_{r}\left(  w\right)
$\ near $z_{0}=z_{m}$. This leads to%
\begin{align}
r\left(  z_{0}\right)   &  \simeq4z_{0}\left[  -2z_{0}g\left(  z_{0}\right)
\frac{\sigma^{\prime}\left(  z_{0}\right)  }{\sigma\left(  z_{0}\right)
}\right]  ^{-\frac{1}{2}},\\
V\left(  z_{0}\right)   &  \simeq4z_{0}\sigma\left(  z_{0}\right)  \left[
-2z_{0}g\left(  z_{0}\right)  \frac{\sigma^{\prime}\left(  z_{0}\right)
}{\sigma\left(  z_{0}\right)  }\right]  ^{-\frac{1}{2}}=\sigma\left(
z_{0}\right)  r\left(  z_{0}\right)  .
\end{align}
From the above expression, we obtain the expected linear potential
$V=\sigma_{s}r$\ at long distance with the string tension,%
\begin{equation}
\sigma_{s}=\left.  \dfrac{dV}{dr}\right\vert _{z_{0}=z_{m}}=\left.
\dfrac{dV/dz_{0}}{dr/dz_{0}}\right\vert _{z_{0}=z_{m}}=\left.  \dfrac
{\sigma^{\prime}\left(  z_{0}\right)  r\left(  z_{0}\right)  +\sigma\left(
z_{0}\right)  r^{\prime}\left(  z_{0}\right)  }{r^{\prime}\left(
z_{0}\right)  }\right\vert _{z_{0}=z_{m}}=\sigma\left(  z_{m}\right)  .
\end{equation}
The temperature dependence of the string tension\ for various chemical
potentials is plotted in figure \ref{string tension}. We see that the string
tension decreases when the temperature increases. At the
confinement-deconfinement transformation temperature $T_{\mu}$, the system
transform to the deconfinement phase and the string tension suddenly drops to
zero as we expected \cite{1006.0055}.

The behaviors of the heavy quark potential at short distance and long distance
agrees with the form of the Cornell potential,%
\begin{equation}
V\left(  r\right)  =-\dfrac{\kappa}{r}+\sigma_{s} r+C,
\end{equation}
which has been measured in great detail in lattice simulations

\begin{figure}[h]
\begin{center}
\includegraphics[
height=2in,
width=3in
]{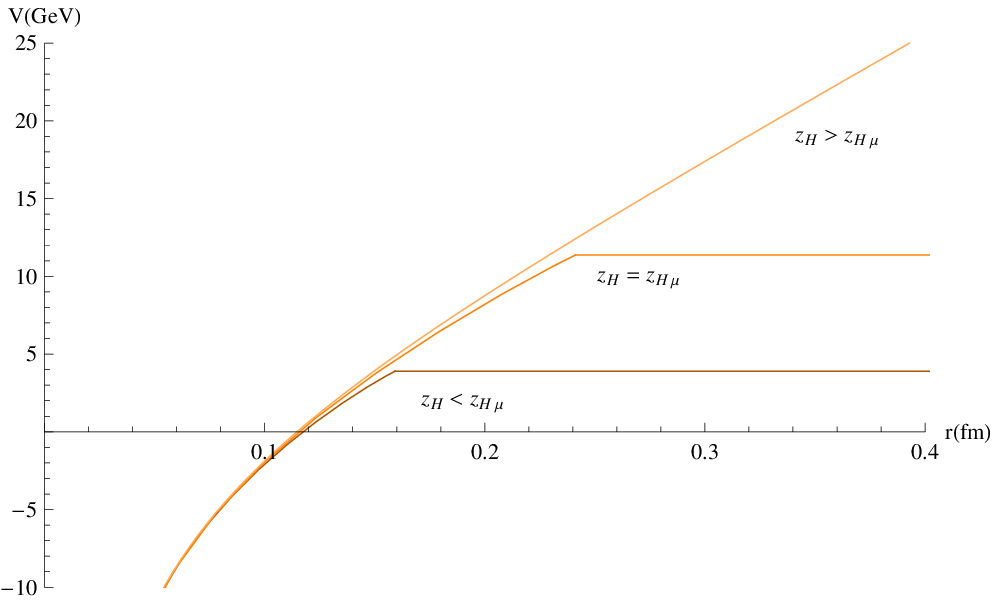}\hspace*{0.5cm} \includegraphics[
height=2in,
width=3in
]{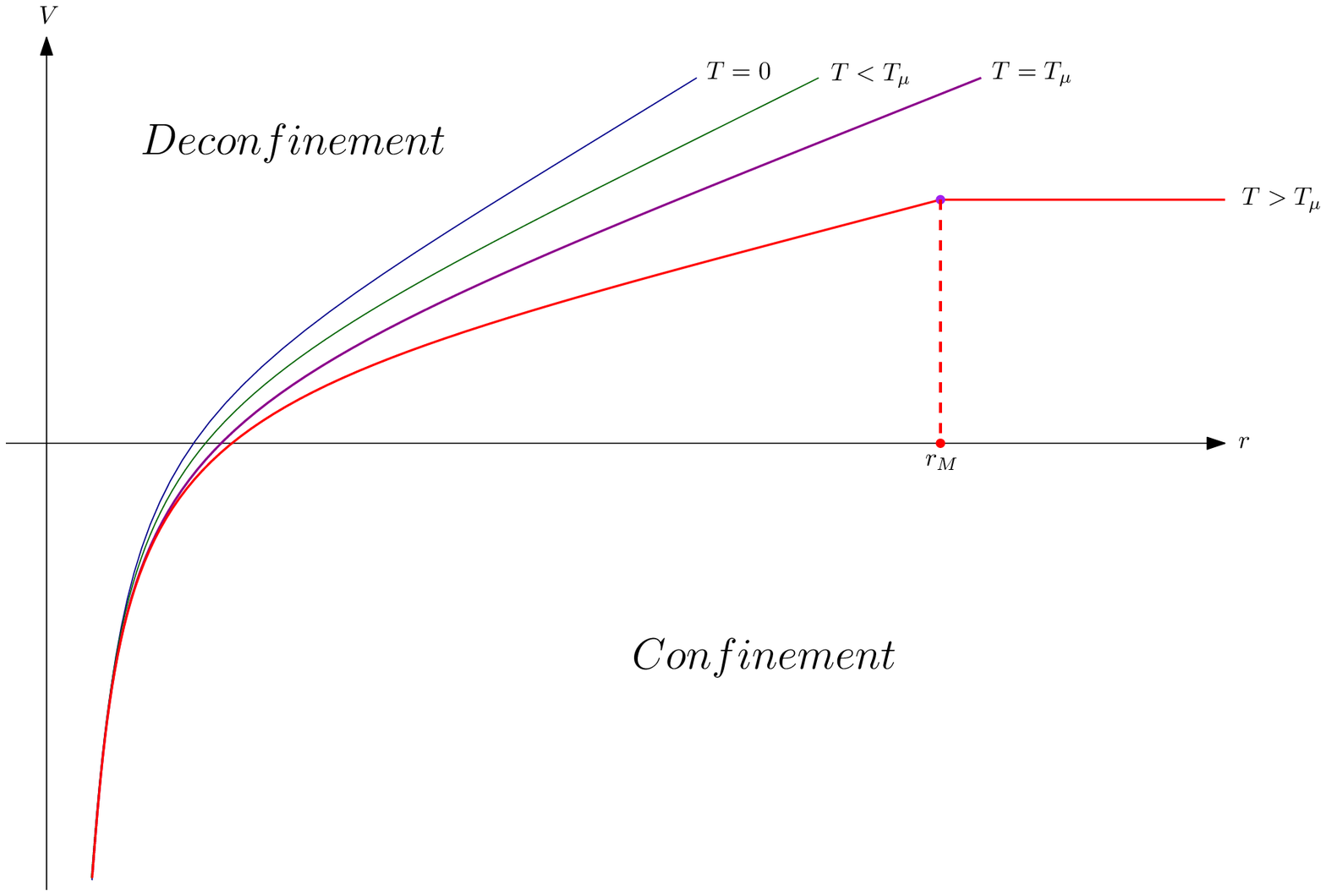}\vskip -0.05cm \hskip 0.15 cm \textbf{( a ) } \hskip 7.5 cm
\textbf{( b )}
\end{center}
\caption{(a) $V$ v.s. $r$ at $\mu=0.5GeV$ for $z_{H}>z_{H\mu}$, $z_{H}%
=z_{H\mu}$ and $z_{H}<z_{H\mu}$. (b) The sketch diagram for heavy quark
potentials at various temperatures at a fixed chemical potential $\mu>\mu_{c}%
$.}%
\label{meson potential}%
\end{figure}

Next, we would like to look at the $r$ dependence of the heavy quark potential
by evaluating the integral in Eq. (\ref{quark potential}), which is divergent
due to its integrand blows up at $z=0$. We simply regularize the integral by
subtracting the divergent part of the integrand,%
\begin{equation}
V_{R}=C\left(  z_{0}\right)  +2\int_{0}^{z_{0}}dz\left[  \frac{\sigma\left(
z\right)  }{\sqrt{g\left(  z\right)  }}\left[  1-\dfrac{\sigma^{2}\left(
z_{0}\right)  }{\sigma^{2}\left(  z\right)  }\right]  ^{-\frac{1}{2}}-\frac
{1}{z^{2}}\left[  1+2A^{\prime}\left(  0\right)  z\right]  \right]  ,
\label{VR}%
\end{equation}
where%
\begin{equation}
C\left(  z_{0}\right)  =-\dfrac{2}{z_{0}}+4A^{\prime}\left(  0\right)  \ln
z_{0}.
\end{equation}
After the regularization, we are able to calculate the heavy quark potential.
The result is plotted in (a) of figure \ref{meson potential}. For short
separate distance, the potential is proportional to $1/r$ as expected. While
for long separate distance, there exists a critical horizon $z_{H\mu}$ for
each chemical potential $\mu$. For small black hole with $z_{H}>z_{H\mu}$, the
potential is linear to $r$ for $r\rightarrow\infty$. While for large black
hole with $z_{H}<z_{H\mu}$, the potential ceases at a maximum distance $r_{M}%
$. Beyond $r_{M}$, the U-shape open string will break to two straight shape
open strings.

It is helpful to use a sketch to describe the heavy quark potential and the
phases transformation with temperature changing. We plot the heavy quark
potentials at various temperatures at a fixed chemical potential\footnote{For
$\mu<\mu_{c}$, the picture is similar but more complicated due to the black
holes phase transition in the background. Here we just illustrate the general
properties for the heavy quark potential and leave the details of the phase
transition to the next section.} $\mu>\mu_{c}$ in (b) of figure
\ref{meson potential}. For $T\leq T_{\mu}$, i.e. $z_{H}>z_{H\mu}$, the heavy
quark potential is linear at large separate distance $r$ with the slopes
decrease with the temperature increasing. The linear potential implies that
the system is in the confinement phase. While for $T>T_{\mu}$, i.e.
$z_{H}<z_{H\mu}$, the heavy quark potential admits a maximum separate distance
$r_{M}$, beyond which the open string breaks up to two straight strings and
the total energy of strings becomes constant. The constant potential implies
that the system is in the deconfinement phase. The confinement-deconfinement
phase transform happens at $T=T_{\mu}$.%

\setcounter{equation}{0}
\renewcommand{\theequation}{\arabic{section}.\arabic{equation}}%

\section{Phase Diagram}

In the previous sections, we have studied the thermodynamics of the black hole
background. We obtained two black hole phases and studied the phase transition
between them. We also added probe open strings in the background and studied
their configurations of U-shape and straight shape, which correspond to
confinement and deconfinement phases in the dual holographic QCD. In
\cite{1301.0385}, we interpreted the black hole to black hole phase transition
in the background as the confinement-deconfinement phase transition in the
dual holographic QCD, leaving a puzzle that a black hole background does not
correspond to the confinement phase in QCD in the original AdS/QCD
correspondence. In this paper, we argued that the U-shape and straight shape
of open strings should correspond to confinement and deconfinement phases in
QCD, but the transformation between the two phases seems always smooth without
phase transition. In this section, by combining these two phenomena, we are
ready to discuss the full phase structure for the system of the open strings
in the black hole background, corresponding to the confinement-deconfinement
phase diagram in the dual holographic QCD.

\subsection{Confinement-deconfinement Phase Diagram}

Let us consider the configurations of the probe open strings first. We have
found that for a small black hole with $z_{H}>z_{H\mu}$, open strings can not
exceed a dynamical wall at $z=z_{m}$\ even the distance $r$ between the
quark-antiquark pair goes to infinity. This means that both ends of the open
strings have to touch the boundary at $z=0$, and the quark-antiquark pair is
always connected by an open string in the U-shape to form a bound state, which
corresponds to a meson state in the dual holographic QCD, as in (a) of figure
\ref{conf-deconf}. We interpret this phase as the confinement phase in QCD. On
the other hand, for a large black hole with $z_{H}<z_{H\mu}$, the two ends of
the open strings could also contact the horizon instead of the boundary. If
the distance $r$ between the quark-antiquark pair is large enough with
$r>r_{M}$, an open string of U-shape\ would break to two straight open strings
as showed in (b) of figure \ref{conf-deconf}. Thus the meson state would decay
to a pair of free quark and antiquark. We interpret this phase as the
deconfinement phase in QCD. We should remark that for a small black hole with
$z_{H}>z_{H\mu}$, it is impossible for an open string to break up due to the
dynamical wall at $z=z_{m}$. Thus even in the black hole background, the
holographic QCD could still be in the confinement phase. This clarifies the
puzzle in \cite{1301.0385} that a black hole background does not correspond to
the confinement phase in QCD in the original AdS/QCD correspondence. The black
hole phases, open string configurations and QCD phases are summarized in Table
\ref{table}.

\begin{table}[h]
\begin{center}%
\begin{tabular}
[c]{|c|c|c|}\hline
Black hole & String configurations for $r\rightarrow\infty$ & Phase in
QCD\\\hline
Small $\left(  z_{H}>z_{H\mu}\right)  $ & U-shape & Confinement\\\hline
Large $\left(  z_{H}<z_{H\mu}\right)  $ & Straight & Deconfenment\\\hline
\end{tabular}
\end{center}
\caption{Black hole phases, open string configurations and QCD phases.}%
\label{table}%
\end{table}

\begin{figure}[h]
\begin{center}
\includegraphics[
height=2in,
width=3in
]{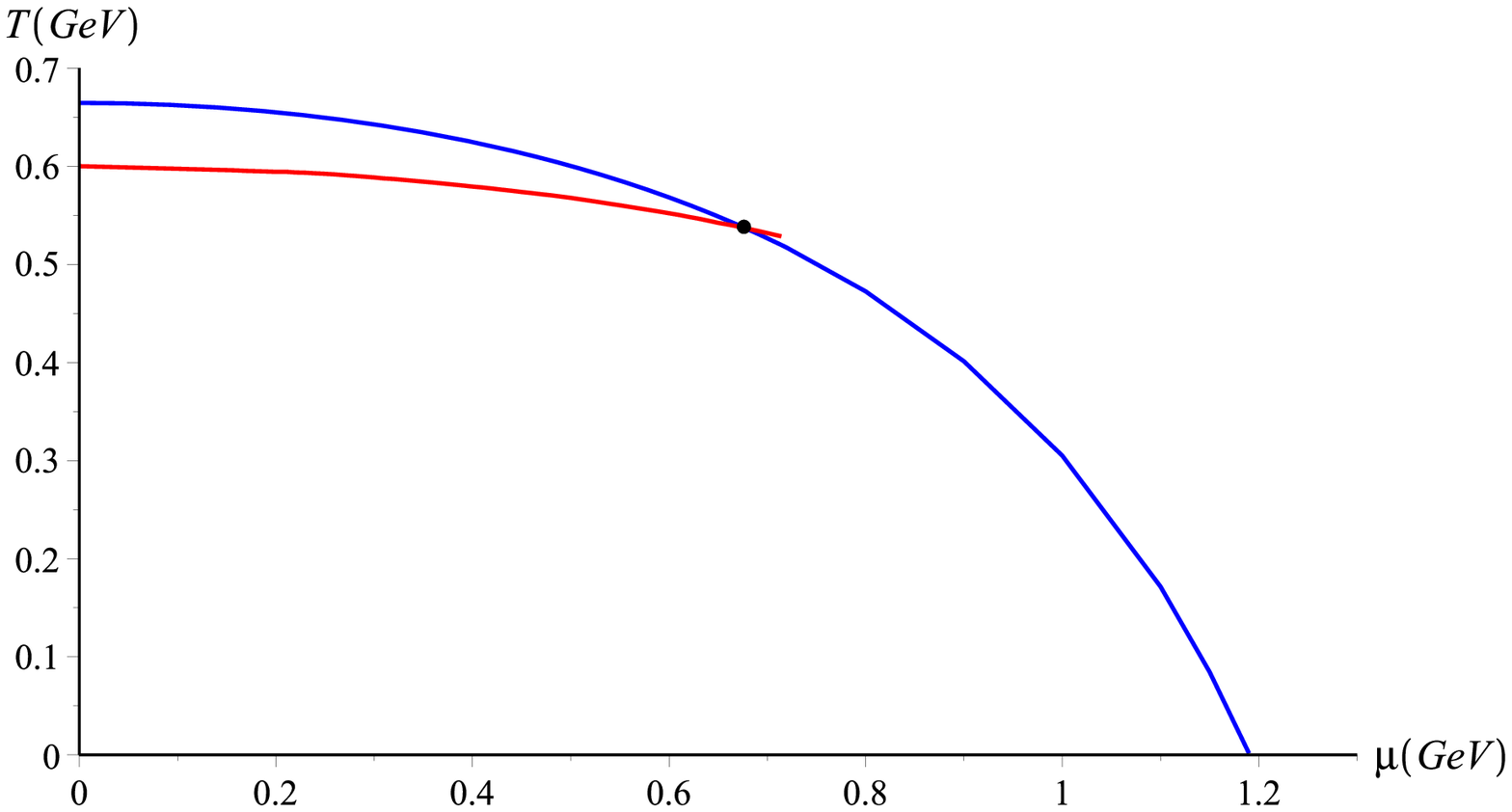}\hspace*{0.5cm} \includegraphics[
height=2in,
width=3in
]{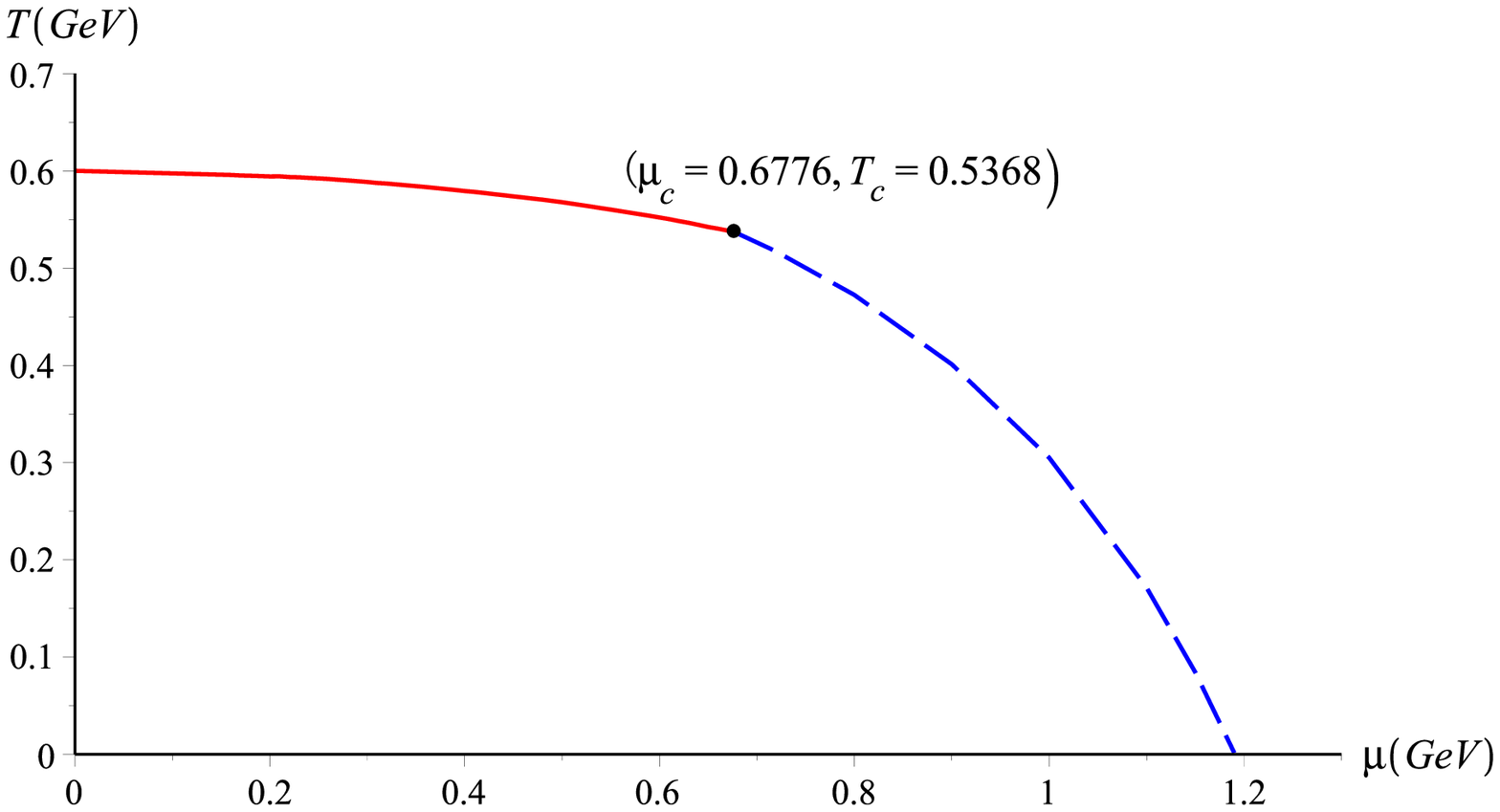}\vskip -0.05cm \hskip 0.15 cm \textbf{( a ) } \hskip 7.5 cm
\textbf{( b )}
\end{center}
\caption{(a) The phase diagrams from the pure black hole background (red line
for $T_{BB}$) and from the configurations of open strings (blue line for
$T_{\mu}$). The two lines intersect at the critical point locates at $\mu
_{c}=0.678GeV$ and $T_{c}=0.536GeV$. (b) The final confinement-deconfinement
phase diagram. For $\mu<\mu_{c}$, there is a phase transition between
confinement and deconfinement phases (red solid line); while for $\mu>\mu_{c}%
$, the phase transition becomes a crossover (blue dashed line).}%
\label{final phase}%
\end{figure}

For each chemical potential $\mu$, we have calculated the transformation
temperature $T_{\mu}$ corresponding to the critical black hole horizon
$z_{H\mu}$. The result of $T_{\mu}$ v.s. $\mu$ is plotted in figure \ref{Tmu}.
On the other hand, the phase transition temperature $T_{BB}$ of black hole to
black hole phase transition in the background was plotted in (b) of figure
\ref{phase diagram}. To investigate the relationship between $T_{\mu}$\ and
$T_{BB}$, we plot both of them together in (a) of figure \ref{final phase}. We
see that the two lines are close to each other but not exactly the same. The
two lines intersect at $(\mu_{c},T_{c})=(0.678GeV,0.536GeV)$, where we define
as the critical point\footnote{We have defined $(\mu_{c},T_{c}%
)=(0.714GeV,0.528GeV)$ as the critical values of the background phase
transition in section 2.2, here we redefined them as the true critical values
of the confinement-deconfinement phase transition.}. For $\mu<\mu_{c}$, when
the temperature increases from zero, the black hole grows with the temperature
and a phase transition eventually happens at $T=T_{BB}(\mu)<T_{\mu}$, where a
small black hole with horizon $z_{Hs}>z_{H\mu}$ suddenly jumps to a large
black hole with horizon $z_{Hl}<z_{H\mu}$ as showed in figure
\ref{phase-cross}. In the dual QCD, this implies that the confinement phase
transform to the deconfinement phase by a phase transition. While for $\mu
>\mu_{c}$, when the temperature increases from zero, the black hole horizon
increases gradually with the temperature and continuously passes $z_{H\mu}$ at
$T=T_{\mu}<T_{BB}(\mu)$. It means that the confinement phase will smoothly
transform to the deconfinement phase as a crossover. Putting everything
together, we obtain the final phase diagram for the confinement-deconfinement
phase transition in QCD, plotted in (b) of figure \ref{final phase}. For the
chemical potential less than the critical point $\mu<\mu_{c}$, we have
confinement-deconfinement phase transition. While for the large chemical
potential $\mu>\mu_{c}$, the confinement-deconfinement phase transition
reduces to a smooth crossover. The critical point is located at $\mu
_{c}=0.678GeV$ and $T_{c}=0.536GeV$. This result is consistent with the
conclusion from the lattice QCD for the heavy quarks \cite{1111.4953}.

\subsection{Meson Melting in Hot Plasma}

\begin{figure}[h]
\begin{center}
\includegraphics[
height=2in,
width=3in
]{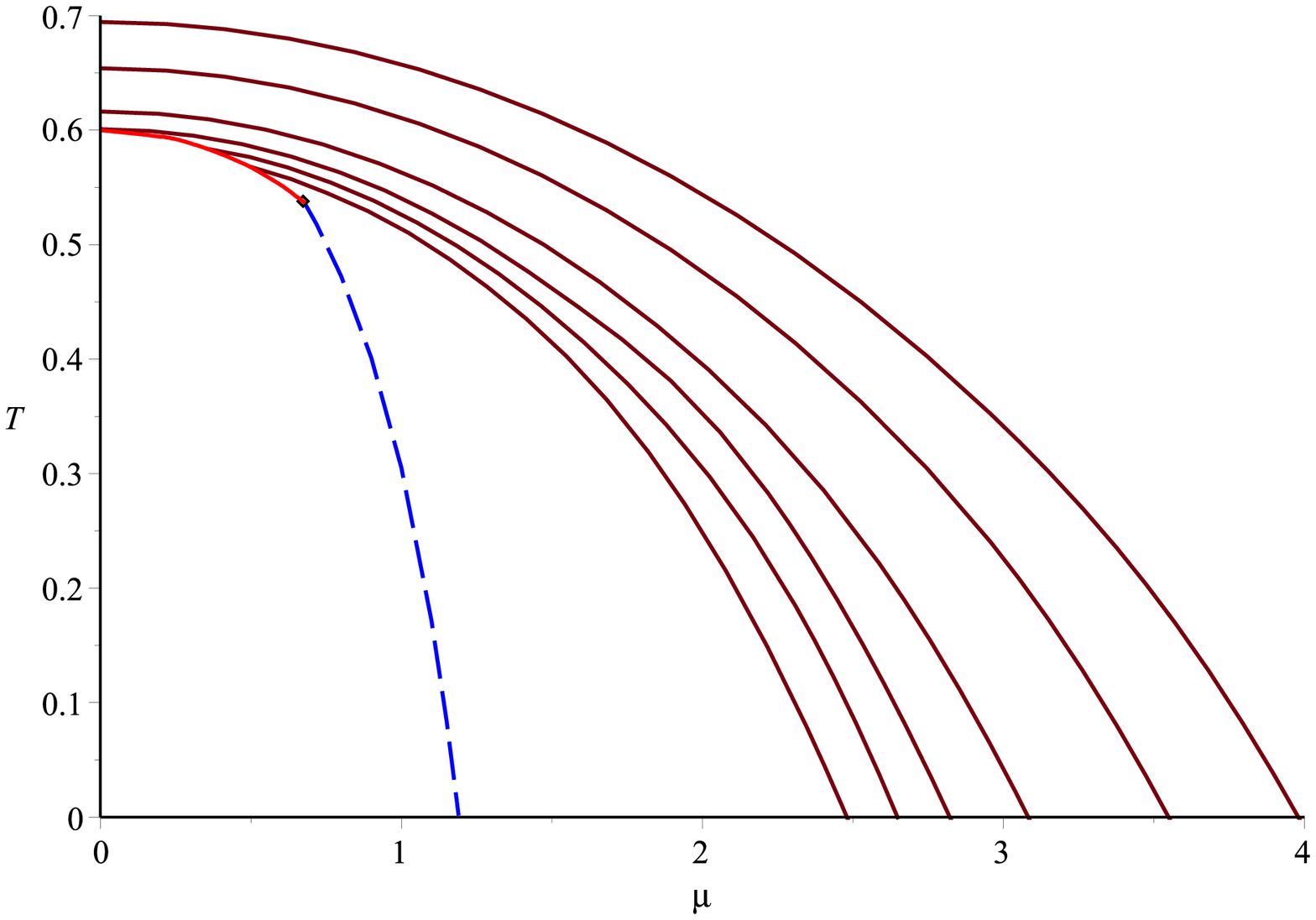}\hspace*{0.5cm} \includegraphics[
height=2in,
width=3in
]{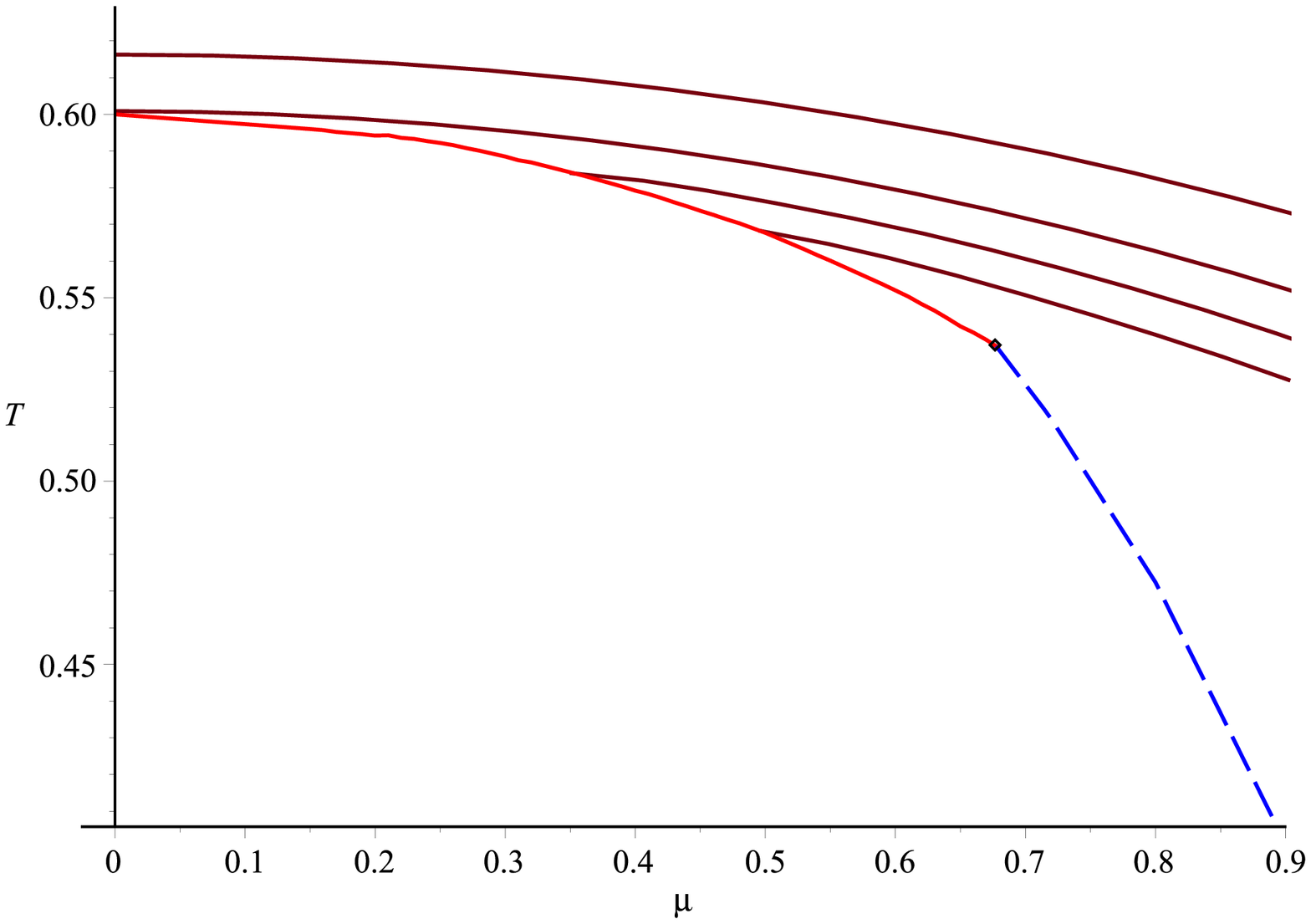}\vskip -0.05cm \hskip 0.15 cm \textbf{( a ) } \hskip 7.5 cm
\textbf{( b )}
\end{center}
\caption{The 'melting lines' for various screening lengths are plotted in (a)
with $r_{M}=0.13,0.15,0.18,0.20,0.22,0.24$ fm from above. In (b), the region
closed to the phase transition line is enlarged.}%
\label{melt}%
\end{figure}

In the confinement phase, open strings are always connected in the
configuration of U-shape, so that the quark-antiquark pair always form a bound
state, i.e. a meson state in QCD. In the deconfinement phase, an open string
could be in the configuration of either U-shape or straight-shape. For short
separate distance $r<r_{M}$, the open string is still in the U-shape. While
for the long separate distance $r>r_{M}$, the energy of the two straight
strings is less than the free energy of an open string in the U-shape and the
U-shape open string will break up to two straight strings, corresponding to
that a meson state melts to a pair of free quark and antiquark.\ The
phenomenon of mesons melting has been previously studied in
\cite{1006.0055,1108.0684,1102.2289,1203.3942}. In this work, we define the
screening length as the maximum length $r_{M}$, achieved by a pair of quark
and antiquark in the bound state at a temperature $T>T_{BB}(\mu)$ for $\mu
\leq\mu_{c}$ or $T>T_{\mu}$ for $\mu>\mu_{c}$. The screening length at a fixed
chemical potential and temperature can be determined by the equation
$V-2F_{q}=0$, where $V$ is the heavy quark potential energy defined in
(\ref{quark potential}) and $F_{q}$ is the free energy of a straight string
defined as%
\begin{equation}
F_{q}=\int_{0}^{z_{H}}dz\frac{e^{2A}}{z^{2}}.
\end{equation}
We plot the 'melting lines' of screening length versus chemical potential in
figure \ref{melt}. The screening length is a possible signal form Quark Gluon
Plasma (QGP). Right after the collision QGP is formed and the temperature is
high enough to be in the deconfinement phase. As temperature decreases, heavy
quarks form bound states at melting temperatures higher than the deconfinement
temperature. This means heavy quark bound sates can coexist with QGP.

\setcounter{equation}{0}
\renewcommand{\theequation}{\arabic{section}.\arabic{equation}}%

\section{Conclusion}

In this paper, we considered an Einstein-Maxwell-scalar system. We solved the
equations of motion to obtain a family of black hole solutions by potential
reconstruction method. We studied the thermodynamical properties of the black
hole background and found black hole to black hole phase transitions at the
temperature $T_{BB}$ for the backgrounds. We then added open strings in these
backgrounds and identified the two ends of an open string as a quark and
antiquark pair in the dual holographic QCD. By solving the equations of motion
of the open strings, we got two configurations for the open strings, i.e.
U-shape and straight-shape. When the temperature is low enough, the black hole
is small, there exists a dynamical wall at $z=z_{m}$ which the open strings
can not exceed even the separation of the quark and antiquark goes to
infinite. From the view of the dual QCD, the quark and antiquark pair is
always connected by an open string to form a bounded state, corresponding to
the confinement phase in QCD. On the other hand, when the temperature is high
enough, the black hole becomes large so that an open string could break up to
two straight open strings connecting the boundary and the black hole horizon,
corresponding to the deconfinement phase in QCD. We obtained the
confinement-deconfinement phase transformation temperature $T_{\mu}$.

Our main conclusion is that, to study the confinement-deconfinement phase
structure in holographic QCD models, we need to combine two phase phenomena in
the bulk gravity theory at the same time, namely the black hole to black hole
phase transition in the background and the various configurations for the
probe open strings. We found that, when the chemical potential is less than
the critical value $\mu_{c}$, the background undergoes a small black hole to a
large black hole phase transition with temperature increasing from zero. The
horizon suddenly blows up to pass the critical horizon $z_{H\mu}$\ so that the
confinement phase transform to the deconfinement phase by a phase transition.
While when the chemical potential is greater than the critical value $\mu_{c}%
$, the black hole horizon grows gradually and continuously pass the critical
horizon $z_{H\mu}$\ so that the confinement phase transform to the
deconfinement phase by a smooth crossover. The final confinement-deconfinement
phase diagram is showed in (b) of figure \ref{final phase}.

We also studied meson melting in this paper. When the temperature is higher
than the phase transition temperature, QCD is in the deconfinement phase.
However, it is known that the meson could still be thermodynamically stable in
the deconfinement phase if the separate distance between the quark and
antiquark is short enough. Only when the separate distance is longer than the
screening length $r_{M}$, the meson becomes unstable and break up to a pair of
free quark and antiquark. We showed the 'melting lines' for various separation
distance in figure \ref{melt}. We conclude that, with increasing temperature,
the mesons of larger size will break up earlier and the mesons of smaller size
will be more stable and break up latter. Inversely, when QGP is cooling down,
the mesons of smaller size will reunite earlier than the ones of larger size.
This will help us to understand the process of the QGP cooling down.

\subsection*{Acknowledgements}

We would like to thank Song He, Mei Huang, Xiao-Ning Wu for useful
discussions. This work is supported by the National Science Council (NSC
101-2112-M-009-005) and National Center for Theoretical Science, Taiwan.

\end{document}